\documentclass[a4paper,11pt]{article}
\pdfoutput=1
\usepackage{jcappub}
\usepackage{bbold}
\usepackage{mathtools}
\usepackage{ulem}
\usepackage{slashed} 
\usepackage[T1]{fontenc}

\allowdisplaybreaks

\renewcommand{\thesection}{\arabic{section}}

\def\laq{~\raise 0.4ex\hbox{$<$}\kern -0.8em\lower 0.62ex\hbox{$\sim$}~}
\def\gaq{~\raise 0.4ex\hbox{$>$}\kern -0.7em\lower 0.62ex\hbox{$\sim$}~}

\def\beq{\begin{equation}}
\def\eeq{\end{equation}}
\def\bea{\begin{eqnarray}}
\def\eea{\end{eqnarray}}

\def \pa {\partial}

\def \Scal {\mathcal{S}}
\def \HH {\mathcal{H}}
\def \ndv {v_{\parallel}}
\def \bn {\mathbf{n}}
\def \bk {\mathbf{k}}
\def \bx {\mathbf{x}}
\def \br {\mathbf{r}}

\title{The skewness of the distance-redshift relation in $\Lambda$CDM}
\author[a,b,c]{T.~Schiavone,}
\author[d]{E.~Di Dio}
\author[a]{and G.~Fanizza}

\affiliation[a]{Instituto de Astrof\'isica e Ci\^encias do Espa\c{c}o, Faculdade de Ci\^encias da Universidade de Lisboa, Edificio C8, Campo Grande, P-1749-016, Lisbon, Portugal}
\affiliation[b]{Department of Physics “E. Fermi”, University of Pisa, Polo Fibonacci, Largo B. Pontecorvo 3, I-56127 Pisa, Italy}
\affiliation[c]{INFN, Istituto Nazionale di Fisica Nucleare, Sezione di Pisa, Polo Fibonacci, Largo B. Pontecorvo 3, I-56127, Pisa, Italy}
\affiliation[d]{CERN, Theory Department, CH-1211 Geneva 23, Switzerland}

\emailAdd{tschiavone@fc.ul.pt}
\emailAdd{enea.didio@cern.ch}
\emailAdd{gfanizza@fc.ul.pt}

\abstract{Starting from a recently proposed framework for the evaluation of the cosmological averages, we evaluate the higher-order moments for the distribution of a given observable. Then, we explicitly discuss the case of the Hubble-Lema\^itre diagram and evaluate its skewness at the leading order in the cosmological perturbative expansion of the gravitational potential. In particular, we focus on perturbations of the luminosity distance due to gravitational lensing. Finally, we discuss our findings in view of recent numerical relativistic simulations, confirming that the skewness in the Hubble-Lema\^itre diagram primarily originates from the late-time matter bispectrum, with other line-of-sight projection effects being sub-dominant.}

\keywords{distance-redshift relation, Hubble-Lema\^itre diagram, non-Gaussianities, matter bispectrum}

\begin{document}

  \begin{minipage}{.45\linewidth}
    \begin{flushleft}
    \end{flushleft}
  \end{minipage}
\begin{minipage}{.45\linewidth}
\begin{flushright}
 {CERN-TH-2023-141}
 \end{flushright}
 \end{minipage}

\maketitle

\section{Introduction}

Luminosity-distance relations have been crucial in establishing the accelerating Universe~\cite{SupernovaSearchTeam:1998fmf,SupernovaCosmologyProject:1998vns} and the standard cosmological model~$\Lambda$CDM. As observational cosmology has progressed, it has moved from being an order-of-magnitude science to one of percent precision. To make effective use of the wealth of unprecedented data now~\cite{BOSS:2016wmc,Planck:2018vyg,DES:2021wwk} and in the coming years~\cite{EUCLID:2011zbd,DESI:2016fyo,LSSTScience:2009jmu,Dore:2014cca,SimonsObservatory:2018koc,Abazajian:2019eic}, it is essential that our theoretical tools evolve accordingly. With regard to the luminosity-distance relation, a key focus is to understand how light propagates in an inhomogeneous Universe and how such effects give rise to observable features. This requires a deeper understanding of the interplay between cosmological structures and the propagation of light, enabling us to identify the subtle signatures imprinted in the observational data.

While the CMB is very well described by linear physics, the study of the Large-Scale-Structure (LSS) of the Universe requires going beyond the linear approximation to extend the range of scales that we can use to constrain cosmological parameters. This can be achieved either by analytical perturbative approaches or non-perturbatively by numerical simulations. In both directions there has been a tremendous improvement in the last decade.
Although numerical simulations can, in principle, handle fully non-linear processes, analytical approaches are computationally much simpler and provide valuable insights into the underlying phenomena. In particular, the two approaches can be compared within the range of validity of perturbation theory.
When describing real observables, it is crucial to consider that discrete sources (galaxies, supernovae, quasars, etc.) are observed through the collection of photons emitted, travelled and detected in a clumpy Universe. Over the last few decades, considerable effort has been devoted to providing a comprehensive relativistic framework for describing LSS within perturbation theory. This began with the pioneering work in the linear approximation~\cite{Yoo:2009au,Challinor:2011bk,Bonvin:2011bg,Jeong:2011as} and then extended to higher orders in perturbation theory~\cite{Yoo:2014sfa,Bertacca:2014dra,DiDio:2014lka,DiDio:2018zmk,DiDio:2020jvo,Magi:2022nfy}. In the meantime, progress has been made in numerical simulations, including full relativistic simulations~\cite{Adamek:2015eda}, as well as the adoption of appropriate gauges to reinterpret Newtonian simulations~\cite{Chisari:2011iq,Fidler:2015npa}.

When interpreting observations, it is necessary to have a robust theoretical framework to describe the statistical properties of the observed structures. This framework must include not only their distribution across the sky (i.e.~the geometric mean), but also their intrinsic stochastic origin, which ultimately shapes the late-time structures. In the last few decades, an active line of research has emerged, aiming at establishing a well-posed mathematical prescription for the averaging of scalar quantities concerning both space-like hypersurfaces \cite{Buchert:1999er,Buchert:2001sa,Buchert:2011sx} and, more recently, generalized also to light-like ones \cite{Gasperini:2009wp,Gasperini:2009mu,Gasperini:2011us,Yoo:2017svj,Fanizza:2019pfp,Buchert:2022zaa}.

The geometry of the Universe, shaped by the distribution of matter, affects the path of photons, leading to shifts in the mean or changes in the probability distribution function (PDF) of observables. In particular, gravitational lensing and the Doppler effect have been studied in relation to the Hubble-Lema\^itre diagram~\cite{Ben-Dayan:2012uam,Ben-Dayan:2012lcv,Ben-Dayan:2013nkf,Umeh:2014ana,Ben-Dayan:2014swa,Bonvin:2015kea,Fanizza:2015swa,Fleury:2016fda,Odderskov:2016rro,Fanizza:2021tuh} and weak lensing measurements \cite{Grimm:2018nto,Yoo:2019qsl,Grimm:2020iyy,Fanizza:2022wob}. 
The non-linear evolution of gravity generates non-Gaussian distributions of matter inhomogeneities, which are subsequently imprinted in the Hubble-Lema\^itre diagram. In addition, non-linear light propagation generates non-Gaussianity in the form of Post-Born effects. These non-Gaussian features have also previously been considered in the context of CMB power spectra~\cite{Pratten:2016dsm,Marozzi:2016uob,Lewis:2016tuj,Liu:2016nfs,Marozzi:2016und,Marozzi:2016qxl,Fabbian:2017wfp} and shear weak lensing~\cite{Schneider:1997ge,Dodelson:2005rf,Petri:2016qya,Bohm:2019bek}. Furthermore, related to fluctuations of the luminosity distance, the non-Gaussianities of the cosmic shear and the local lensing convergence induced by source clustering were investigated in Refs.~\cite{Bernardeau:1997tj,Schneider:1997ge,VanWaerbeke:2000ed,Hamana:2000wb}.

Non-Gaussian distributions are characterised by non-vanishing higher moments. In our work, we introduce the formalism for evaluating the skewness of any cosmological observable on the light cone at the leading order in perturbation theory. In applying this formalism to the luminosity distance we are motivated by the numerical results presented in Ref.~\cite{Adamek:2018rru}, which highlighted significant deviations from Gaussianity in the luminosity distance PDF. We will show that the leading order terms solely involves second-order perturbations of the luminosity distance. However, when dealing with 1-point functions, a direct comparison is not possible due to the limitations within the validity of perturbation theory. To remove the non-linear effects, small scales must be smoothed out in the analysis. Another key question is whether these non-Gaussian features are a direct evidence of the expected late-time non-Gaussianities or they are somehow sourced by spurious effects in the simulations, such as the finite sampling, limited sky-coverage or redshift binning.

Despite these limitations, our results are in line with those obtained from the numerical light tracing in Ref.~\cite{Adamek:2018rru}, especially as we approach higher redshifts, where the effect of non-linearities diminishes at fixed scale. While a more direct comparison, where the numerical and perturbative approaches are smoothed at the same physical scale, remains a prospect for future work, our study confirms that the dominant contribution to the skewness comes from the matter bispectrum, which affects the lensing propagation. Meanwhile, other relativistic effects play a sub-dominant role in the observed skewness.

In summary, in this paper, motivated by the results obtained in \cite{Adamek:2018rru}, we are interested in characterizing the distribution of the luminosity distance in order to quantify non-Gaussianities. Our work aims to develop a general method to evaluate analytically non-Gaussianities for any desired cosmological observables. Then, we will apply our general results to the Hubble-Lema\^itre diagram. In this regard, bearing in mind the general average prescription in cosmology over the past light-cone \cite{Fanizza:2019pfp}, we will focus on the skewness of the distribution of a generic scalar observable, like the luminosity distance-redshift relation, averaged over a region of space-time.

This paper is organised as follows. In Sect.~\ref{Sec1}, we review the general features of the evaluation of the average and dispersion at the leading order in perturbation theory. In Sect.~\ref{highorder}, we apply the same formalism for the evaluation of the average and dispersion to the computation of the skewness at the leading order. Sect.~\ref{sec:hodz} is devoted to the explicit evaluation of the skewness for the distance-redshift relation in the perturbed FLRW Universe. Here, we limit our computation to the lensing terms. Afterwards, we specialise our science case to the number count weighted measure in the infinitesimal redshift bin limit. In Sect.~\ref{sec:analytic}, we explicitly evaluate the leading terms to the skewness and show how they are connected to the matter power spectrum and to the matter bispectrum. Numerical results for a fiducial $\Lambda$CDM model are presented in Sect.~\ref{sec:numerics}, whereas Sect.~\ref{sec:num_sim} is devoted to the (possible) comparison with numerical simulations. Finally, our main conclusions are outlined in Sect.~\ref{sec:conc}. Appendices~\ref{app:hopert} and \ref{app:analytic} report explicit technical evaluations needed to derive our results.

In this work we adopt the following 3-D Fourier transform convention 
\begin{equation}
f \left(\eta, \bx \right) = \int \frac{d^3k}{\left( 2 \pi \right)^3} \ f \left( \eta,\bk \right) \ e^{-i \bx \cdot \bk}\qquad, \qquad
f \left( \eta,\bk \right) = \int d^3x \ f \left( \eta,\bx \right) \ e^{i \bx \cdot \bk} \,.
\label{eq:Fourier_conv}
\end{equation}

\section{Leading-order terms review: average and dispersion}
\label{Sec1}

In this section, we will discuss how to evaluate the moments of a given observable. The specific choice for the measure adopted for their evaluation will be left for later. In particular, we will adopt the covariant prescriptions for different kinds of light-cone averages presented in \cite{Fanizza:2019pfp}.

Let us start then by considering a generic observable $\Scal$. Its average over a given portion of spacetime is given, in complete generality, by
\bea
\langle \Scal \rangle = \frac{\int d\mu\,\Scal}{\int d\mu}\,,
\label{eq:start}
\eea
where $\mu$ represents the measure weighting the average procedure. Eq.~\eqref{eq:start} is exact but, for our purposes, can be expanded perturbatively.
We work in a perturbed FLRW metric for scalar perturbations
\begin{equation}
ds^2=a^2(\eta)
\left\{ -\left( 1+2\Phi \right)d\eta^2 +(1-2\Psi)\left[ dr^2+r^2 \left(d\theta^2 + \sin^2\theta d\varphi^2\right) \right] \right\}\,
\end{equation}
where\footnote{For the sake of simplicity, we will omit the index ${}^{(1)}$ for linear perturbations of the gravitational potentials.} $\Phi\equiv \sum_i \phi^{(i)}$ and $\Psi\equiv \sum_i \psi^{(i)}$ and $i$ refers to the order of the perturbative expansion. So by expanding to second order the observable $\Scal$ and the measure $\mu$, we have
\bea
\Scal\simeq&\,\Scal^{(0)}\left(1+\sigma^{(1)}+\sigma^{(2)}\right)\,,
\nonumber\\
d\mu\simeq&\,d\mu^{(0)}\left( 1+\mu^{(1)}+\mu^{(2)}\right)\, .
\label{eq:pert}
\eea
The fundamental requirement to apply Eq.~\eqref{eq:start} is that $\Scal$ transforms as a  scalar field under a generic coordinate transformation. With this proviso, Eq.~\eqref{eq:start} can be applied to the generic $\alpha$-th power of $\Scal$ as well.

According to the perturbative scheme outlined in Eqs.~\eqref{eq:pert}, we can then rescale our observable to its background value, namely apply Eq.~\eqref{eq:start} to $\Scal/\Scal^{(0)}$ rather than $\Scal$. On one hand, this choice simplifies a lot the equations and is in line with previous estimations provided in literature (see \cite{Ben-Dayan:2012uam}). On the other hand, and more importantly, this is the result that has been provided by numerical simulation (see, for instance, the histogram in Fig. 2 of \cite{Adamek:2018rru}). We then have
\bea \label{eq:avpower}
\left\langle \left(\frac{\Scal}{\Scal^{(0)}}\right)^\alpha\right\rangle&\simeq&\frac{\int d\mu^{(0)}\,\left[1
+\mu^{(1)}
+\alpha\,\sigma^{(1)}
+\mu^{(2)}
+\alpha\,\mu^{(1)}\sigma^{(1)}
+\frac{\alpha}{2}\left( 2\,\sigma^{(2)}+\left( \alpha - 1 \right){\sigma^{(1)}}^2 \right)\right]}{\int d\mu^{(0)}\,(1+\mu^{(1)}+\mu^{(2)})}\nonumber\\
&\simeq&1
+\alpha\,I[\sigma^{(1)}]
\nonumber\\
&&
+\alpha\,I[\mu^{(1)}\,\sigma^{(1)}]
+\alpha\,I[\sigma^{(2)}]
+\frac{\alpha\,\left( \alpha - 1 \right)}{2}\,I\left[{\sigma^{(1)}}^2\right]
-\alpha\,I[\mu^{(1)}]\,I[\sigma^{(1)}]\,,
\eea
where we have defined
\beq
I[f]\equiv\frac{\int d\mu^{(0)} f}{\int d\mu^{(0)}}\,.
\label{eq:definition_I[f]}
\eeq
The average meant by $\langle \dots\rangle$ is a geometrical procedure which takes into the smoothing of an arbitrary inhomogeneous manifold but does not specify the nature of these inhomogeneities. This nature is given by a further ensemble average $\overline{{\color{white}a}\dots{\color{white}a}}$ which provides the stochastic properties of the inhomogeneities. Having in mind that our perturbations will be sourced by the fluctuations of the gravitational potentials $\Phi$ and $\Psi$, we just assume for now that for the linear gravitational potentials $\overline{\phi}=\overline{\psi}=0$ and that higher-order powers of $\Phi$ and $\Psi$ can have a non-vanishining ensemble average. We will provide later the mathematical properties for these operations. We then get
\bea \label{eq:avpower_ensemble}
\overline{\left\langle \left(\frac{\Scal}{\Scal^{(0)}}\right)^\alpha\right\rangle}
&=&1
+\alpha\,\overline{I[\mu^{(1)}\,\sigma^{(1)}]}
+\alpha\,\overline{I[\sigma^{(2)}]}
+\frac{\alpha\,\left( \alpha - 1 \right)}{2}\,\overline{I\left[{\sigma^{(1)}}^2\right]}
-\alpha\,\overline{I[\mu^{(1)}]\,I[\sigma^{(1)}]}\,,
\eea
where we used the abovementioned properties that the ensemble average of linear perturbations vanishes.

Eq.~\eqref{eq:avpower_ensemble} can be readily applied to the evaluation of the average of $\Scal/\Scal^{(0)}$: for $\alpha =1$ indeed we have
\bea \label{eq:msecond}
m \equiv\overline{\left\langle \frac{\Scal}{\Scal^{(0)}} \right\rangle}
&=&1
+\overline{I[\mu^{(1)}\,\sigma^{(1)}]}
+\overline{I[\sigma^{(2)}]}
-\overline{I[\mu^{(1)}]\,I[\sigma^{(1)}]}\,.
\eea
This result already provides an important intermediate step to evaluate the generic $\alpha$-th moment $\mu_\alpha$ of the distribution of $\Scal$. In fact, this can be written as
\bea \label{eq:mu_alpha}
\mu_\alpha & \equiv &\overline{\left\langle\left(\frac{\Scal}{\Scal^{(0)}}-m\right)^\alpha\right\rangle}
=\sum_{k=0}^\alpha(-1)^{\alpha-k}\binom{\alpha}{k}m^{\alpha-k}\overline{\left\langle\left(\frac{\Scal}{\Scal^{(0)}}\right)^k\right\rangle}\nonumber\\
&=&\sum_{k=0}^\alpha(-1)^{\alpha-k}\binom{\alpha}{k}
\left\{1
+\alpha\,\overline{I[\mu^{(1)}\,\sigma^{(1)}]}\right.\nonumber\\
&&\left.+\alpha\,\overline{I[\sigma^{(2)}]}
-\alpha\,\overline{I[\mu^{(1)}]\,I[\sigma^{(1)}]}
+\frac{k\,\left( k - 1 \right)}{2}\,\overline{I[{\sigma^{(1)}}^2]}
\right\}\,.
\eea
Eq.~\eqref{eq:mu_alpha} is useful also for the evaluation of the variance, i.e. $\sigma^2\equiv \mu_2$, where we have
\begin{equation}\label{eq:variance}
\sigma^2=\overline{I[{\sigma^{(1)}}^2]}\,.
\end{equation}
However, all the higher moments $\mu_\alpha$ with $\alpha>2$ vanish. This is due to the fact that in this section we have limited our non-linearities to be at most of second-order in perturbation theory. In the next section, we will show how the next-to-leading order terms will impact the analytic estimation of higher-order multipoles, such as $\mu_3$.

\section{Next-to-leading order term: skewness}
\label{highorder}

In this section, we will provide the analytic evaluation for the skewness
of a generic observable $\Scal$. As said in the previous section, we recall that we remove the background contribution for the observable by studying $\Scal/\Scal^{(0)}$ rather than just $\Scal$. In order to proceed to our end, we first need to introduce, in complete generality, the standardized moments $\kappa_\alpha$ as
\begin{equation}\label{eq:standard}
\kappa_\alpha\equiv \frac{\mu_\alpha}{\left(\sigma^2\right)^{\alpha/2}}
=\frac{1}{\left(\sigma^2\right)^{\alpha/2}}
\overline{\left\langle \left( \frac{\Scal}{\Scal^{(0)}}-m \right)^\alpha \right\rangle}\,.
\end{equation}
However, as we have pointed out in the previous section, second-order corrections are not enough to evaluate higher-order moments. Naively, one would expect that each $\mu_\alpha$ demands $\alpha$-th order perturbations to get the leading term. This estimation is accurate when $\alpha$ label even moments. For what concerns the odd moments, however, leading terms are of $\alpha+1$-th order in perturbation theory (as an instance, this occurs for the average and the dispersion). This peculiar hierarchy is due to the fact the leading term for each moment $\mu_\alpha$ is $\overline{I[{\sigma^{(1)}}^\alpha]}$. Therefore, since $\sigma^{(1)}\sim \psi^{(1)}$, we have that the $\alpha$-th order term of $\mu_\alpha$ is related to the $\alpha$-point correlation function of $\psi^{(1)}$: as a consequence of Wick theorem, by assuming Gaussian initial conditions, the latter is non-null only when $\alpha$ is even.

According to what discussed so far, the evaluation of higher order moments would require the estimation of perturbations for both the measure $\mu$ and the observable $\Scal$ beyond the second order at least for the leading terms. These quantities are not available in literature and evaluating them is a non-trivial task which would be an interesting result \textit{per se}. Fortunately, it can be shown that the expansion in Eqs.~\eqref{eq:pert} still catches all terms for what concerns third moment at leading order even when perturbations of $\Scal$ and $d\mu$ up to the fourth order are consistently taken into account. The derivation is quite long but straightforward and the interested reader can refer to Appendix \ref{app:hopert} for this proof. Hence, we get
\bea \label{eq:skewness}
\mu_3
&=&\overline{I[\sigma^{(1)\,3}]}
+\overline{I[ \sigma^{(1)\,3}\mu^{(1)}]}
-\overline{I[ \sigma^{(1)\,3}]I[\mu^{(1)}]}
+3\overline{I[\sigma^{(1)\,2} \sigma^{(2)}]}\nonumber\\
&&+3\,\overline{I[ \sigma^{(1)\,2}]}\,\,\overline{I[\mu^{(1)}]\,I[\sigma^{(1)}]}
-3\overline{I[\sigma^{(1)\,2}]}\,\,\overline{I[\mu^{(1)}\,\sigma^{(1)}]}
-3\overline{I[\sigma^{(1)\,2}]}\,\,\overline{I[\sigma^{(2)}]}\nonumber\\
&=&\overline{I[ \sigma^{(1)\,3}\mu^{(1)}]}
+3\overline{I[\sigma^{(1)\,2} \sigma^{(2)}]}
-\overline{I[ \sigma^{(1)\,3}]I[\mu^{(1)}]}
-3\,\sigma^2\,\left(m-1\right)\,.
\eea
where $m$ and $\sigma^2$ are respectively given by Eqs.~\eqref{eq:msecond} and \eqref{eq:variance}. We remark again the disappearing of first term in the first line of Eq.~\eqref{eq:skewness}: this would be the only third-order contribution but vanishes as a consequence of Wick theorem, as anticipated before. This result is completely general in regard of the kind of observable $\Scal$ and the chosen prescription of the spatial average $\mu$. In the next section, we will apply them to the specific cases of the distance-redshift relations.

\section{Higher-order moments for the distance-redshift relations}
\label{sec:hodz}
In this section, we want to apply the general formalism previously developed to the explicit case of the luminosity distance-redshift relation 
$d_L(z)$. 
As recently pointed out in Ref.~\cite{Adamek:2018rru}, the probability distribution of the luminosity distance $d_L(z)$ exhibits a significant non-Gaussian contribution. Our objective is to investigate the extent to which this behavior can be captured within perturbation theory.

Linear and non-linear perturbations for the distance-redshift relation have been discussed in several papers \cite{Sasaki:1987ad,Bonvin:2005ps,Ben-Dayan:2012lcv,Umeh:2014ana,Biern:2016kys,Koksbang:2018iuh}. For the purposes of this work, we refer to the leading lensing terms as given in Eqs.~(B.1) and (B.2) of \cite{Fanizza:2015swa}. Moreover, we will assume that there is no anisotropic stress at linear order, i.e.~$\phi=\psi$. 
Then, we can express the luminosity distance to second order in perturbation theory as follows\footnote{From now on, unless otherwise specified, $\sigma^{(1)}$ and $\sigma^{(2)}$ refer to the perturbations of the luminosity distance-redshift relation.}
\begin{equation}
    d_L \left( z \right) \simeq d^{(0)}_L\left( z \right) \left( 1 + \sigma^{(1)} + \sigma^{(2)} \right)\,,
\end{equation}
with
\bea
\sigma^{(1)}&=&
\int_0^{r_s} dr \frac{r-r_s}{r r_s} \Delta_2 \psi \left( r \right)\,,
\nonumber\\
\sigma^{(2)}
&=&\frac{1}{2}\sigma^{(1)\,2}+\Sigma^{(2)}+\sigma^{(2)}_{LSS}\,,
\label{eq:52}
\eea
and where $d^{(0)}_L\left( z \right)$ is the luminosity distance in a  FLRW background, $r_s$ is the comoving distance to the source, $\Delta_2$ is the dimensionless angular Laplacian on the 2-sphere, and $\psi(r)=\psi(\eta_o-r,r)$. We have also introduced\footnote{As a remark, here we have followed the same convention for the second order gravitational potential as \cite{DiDio:2015bua}, differently from \cite{Fanizza:2015swa}, where $\Phi\equiv\phi+\frac{1}{2}\phi^{(2)}$ and $\Psi\equiv\psi+\frac{1}{2}\psi^{(2)}$. This will lead to a difference in the prefactor of $\sigma^{(2)}_{LSS}$ later on.} 
\bea
\sigma^{(2)}_{LSS}&\equiv&\frac{1}{2}\int_0^{r_s}\,dr\frac{r-r_s}{rr_s}
\Delta_2\left[ \psi^{(2)}+ \phi^{(2)} \right](r)\,,
\nonumber\\
\Sigma^{(2)}&\equiv&2\int_0^{r_s}\,dr\frac{r-r_s}{rr_s}\pa_b\left[ \Delta_2\psi(r) \right]
\int_0^{r_s}dr\,\frac{r-r_s}{rr_s}\bar\gamma_0^{ab}\pa_a\psi(r)\nonumber\\
&&+2\,\int_0^{r_s}dr\left\{ \gamma_0^{ab}\pa_b\left[ \int_0^rdr'\,\psi(r') \right]
\int_0^rdr'\,\frac{r'- r}{rr'}\pa_a\Delta_2\psi(r') \right\}\nonumber\\
&&+\int_0^{r_s}dr\,\frac{r-r_s}{rr_s}\Delta_2\left[ \gamma_0^{ab}\pa_a\left( \int_0^rdr'\,\psi(r') \right)\pa_b\left( \int_0^rdr'\,\psi(r') \right) \right]\,,
\label{eq:53}
\eea
where $\bar\gamma_0^{ab}=\text{diag}(1,\sin^{-2}\theta)$, and $\gamma_0^{ab}=r^{-2}\bar\gamma_0^{ab}$.

In order to evaluate the skewness of the PDF, we also need to choose a measure $\mu$, according to Eq.~\eqref{eq:skewness}. Several possibilities have been discussed in \cite{Fanizza:2019pfp} for measures, which are general covariant and gauge invariant. In this work, we have chosen to utilize the galaxy number count weighted measure for computing the averages. This measure has been discussed for the backreaction of stochastic inhomogeneities to the mean value of the Hubble diagram in the limit case of small width of the redshift bin \cite{Yoo:2017svj,Fleury:2016fda,Fanizza:2019pfp}.
However, it can be applied directly also for the case of finite redshift bin, preserving the property of general covariance \cite{Fanizza:2019pfp}.

In regard to the comparison with numerical simulations, the adoption of the average weighted with galaxy number count appears to be the most appropriate choice. This weighting scheme naturally assigns more significance to regions with higher density contrasts, allowing for a more accurate and meaningful comparison.
The background term in the measure is then given by
\begin{equation}\label{eq:back_measure}
d\mu^{(0)}=-\left[\frac{\rho(z)\,d^2_A(z)}{(1+z)H(z)}\right]^{(0)}dz\,d\Omega\,,
\end{equation}
where $d\Omega$ is the infinitesimal solid angle, $d^{(0)}_A$ is the background angular distance, $\rho^{(0)}$ is the background density and the denominator comes from the background expansion of the number counts. We are then left with reporting the linear order galaxy number counts~\cite{Yoo:2009au,Yoo:2010ni,Challinor:2011bk,Bonvin:2011bg}. The leading terms in the weak field expansion are given by \cite{DiDio:2014lka}
\beq \label{eq:linear_number_count}
\mu^{(1)}=\delta
+ \HH^{-1} \partial_r \ndv
+2\int_0^{r_s}\,dr\frac{r-r_s}{r r_s}\Delta_2\psi(r)
=\delta
+ \HH^{-1} \partial_r \ndv
+2\,\sigma^{(1)}\,,
\eeq
where $\delta$ is the matter linear perturbation (with the galaxy bias set to $b_1=1$) and $\HH^{-1}\pa_r v_\rVert$ accounts for the redshift space distortion.

With Eqs.~\eqref{eq:back_measure} and \eqref{eq:linear_number_count}, we can compute $\mu_3$.
In details, from Eq.~\eqref{eq:skewness}, we explicitly get
\bea \label{eq:skewness_number_count}
\mu_3
&=&\frac{7}{2}\,\overline{I[ \sigma^{(1)\,4}]}
+3\overline{I\left[\sigma^{(1)\,2} \sigma^{(2)}_{LSS}\right]}
+3\overline{I[\sigma^{(1)\,2} \Sigma^{(2)}]}
\nonumber\\
&&
+\overline{I[ \sigma^{(1)\,3}\delta]}
+\overline{I\left[ \frac{\sigma^{(1)\,3}\pa_r v_\rVert}{\HH}\right]}
-3\,\sigma^2\,\left(m-1\right)\,,
\eea
where leading order term for $\sigma^2$ is given by Eq.~\eqref{eq:variance}, and the average is given by
\beq \label{eq:msecond_number_count}
m-1
=\overline{I[\sigma^{(1)}\delta]}
+\overline{I\left[\frac{\sigma^{(1)}\pa_r v_\rVert}{\HH}\right]}
+\frac{5}{2}\,\overline{I[\sigma^{(1)\,2}]}
+\overline{I[\Sigma^{(2)}]}\,.
\eeq
In the derivation of Eq.~\eqref{eq:skewness_number_count} we made use of $I[\sigma^{(1)}]=0$. This is because we have considered only lensing terms in Eqs.~\eqref{eq:52} and terms sourced by a Laplacian vanish when averaged over spatial directions. The same result holds also in the derivation of Eq.~\eqref{eq:msecond_number_count} in regard of the term $I\left[\sigma^{(2)}_{LSS}\right]$. In a similar manner, also terms as $\overline{I[ \sigma^{(1)\,3}]I[\delta]}$ and $\overline{I[ \sigma^{(1)\,3}]I\left[\HH^{-1} \pa_r v_\rVert\right]}$ do not contribute to Eq.~\eqref{eq:skewness_number_count}. This is because the ensemble average for these terms always correlates the monopole of $\delta$ or $\HH^{-1}\pa_r v_\rVert$ with the Laplacian of $\sigma^{(1)}$, selecting then its null eigenvalue. Sub-leading relativistic corrections to the distance-redshift relation, such as Doppler effect, can contribute with a non-vanishing monopole, especially if sources at small redshifts are taken into account. However, for the purposes of this pioneering work, we have not considered them, even thought they can be of interest for the analysis of close sources. We will investigate their presence in a forthcoming work.

\subsection{Infinitesimal redshift bin}\label{sec:infinitesimal}
Before concluding this section, some comments about the integration domain are in order. The integration over the solid angle $d\Omega$ in Eq.~\eqref{eq:back_measure} covers the entire solid angle. However, for the integration over the redshift we adopt a redshift bin of width $\Delta z$ such that all the needed averages are evaluated within a certain redshift range $[z_s,z_s+\Delta z]$. In case of large value of the redshift bin width, all the redshift dependent terms in Eq.~\eqref{eq:back_measure} needs to be integrated. However, in the limit when $\Delta z$ becomes small enough to be considered infinitesimal\footnote{This limit has been originally proposed in \cite{Fleury:2016fda} and its covariance has been then proved in \cite{Fanizza:2019pfp}.}, i.e.~$\Delta z\rightarrow \delta z$, the measure in Eq.~\eqref{eq:back_measure} simplifies to
\begin{equation}\label{eq:back_measure_inf}
d\mu^{(0)}\simeq-\left[\rho(z_s)\,d^2_A(z_s)\right]^{(0)}\frac{\delta z\,d\Omega}{(1+z_s)H(z_s)}\,,
\end{equation}
and then the integrals in the definition of $I[f]$ in Eq.~\eqref{eq:definition_I[f]} reduce to simple angular integrals. Moreover, within this limit, all the redshift dependent quantities and $\delta z$ itself factorize out of the integrals and cancel out in the ratio with the unconnected diagram in Eq.~\eqref{eq:definition_I[f]}, returning then
\begin{equation}
\lim_{\Delta z\rightarrow \delta z}I[f] = \frac{1}{4\pi}\int d\Omega f(z_s,{\bf n})
\label{eq:limit}\,.
\end{equation}

In this limit, the non-purely second-order terms in the first equality of Eq.~\eqref{eq:skewness} cancel. To this end, we anticipate in the limit of Eq.~\eqref{eq:limit} that
\bea
\overline{I[ \sigma^{(1)\,3}\mu^{(1)}]}&=&
3\,I\left[\overline{\sigma^{(1)\,2}}\,\overline{\mu^{(1)}\sigma^{(1)}}\right]
=
3\,\overline{\sigma^{(1)\,2}}\,\overline{\mu^{(1)}\sigma^{(1)}}
\nonumber\\
&=&3\,\overline{I[\sigma^{(1)2}]}\,\overline{I[\mu^{(1)}\sigma^{(1)}]}
=3\,\sigma^2\overline{I[\mu^{(1)}\sigma^{(1)}]}\,,
\label{eq:simplify-infinitesimal-bin}
\eea
where  we have used Eq.~\eqref{eq:variance} and the fact that the ensemble average can not have any preferred direction due to statistical isotropy. This term exactly cancels the counterpart coming from Eq.~\eqref{eq:msecond} in Eq.~\eqref{eq:skewness_number_count}. Here, we make the first interesting remark that the weight for the measure is irrelevant in the infinitesimal bin limit for the leading order terms of the skewness for the luminosity distance-redshift case study. Moreover, in this limit, the third moment reduces to the simpler expression
\bea
\mu_3
&=&\mu^Q_3+\mu^{PB}_3+\mu^{LSS}_3\,,
\label{eq:leading_skewness}
\eea
where we have defined
\bea
\mu^Q_3&\equiv&\frac{7}{2}\,\overline{I[ \sigma^{(1)\,4}]}
-\frac{15}{2}\,\left(\sigma^2\right)^2\,,
\nonumber\\
\mu^{PB}_3&\equiv&3\left\{\overline{I[\sigma^{(1)\,2} \Sigma^{(2)}]}
-\sigma^2\overline{I[\Sigma^{(2)}]}\right\}\,,
\nonumber\\
\mu^{LSS}_3&\equiv&3\,\overline{I\left[\sigma^{(1)\,2} \sigma^{(2)}_{LSS}\right]}\,.
\label{eq:skewness_label}
\eea
We have combined Eqs.~\eqref{eq:skewness_number_count} and \eqref{eq:msecond_number_count}, adapted for an infinitesimal redshift bin by using Eq.~\eqref{eq:simplify-infinitesimal-bin}, to cancel out terms like $\overline{I[ \sigma^{(1)\,3}\delta]}$ and $\overline{I\left[ \frac{\sigma^{(1)\,3}\pa_r v_\rVert}{\HH}\right]}$, as previously mentioned.

Eq.~\eqref{eq:leading_skewness} is entirely sourced by pure non-linear terms in the expression of the distance-redshift relation. 
Indeed, when considering Gaussian initial conditions, linearly evolved perturbations preserve their probability distribution. Consequently, they do not generate any non-Gaussianity in the limit of infinitesimally narrow bins, where the hypersurfaces for the averages are evaluated at constant redshift.

Hence, the terms in Eq.~\eqref{eq:leading_skewness} are sourced by pure non-Gaussian effects and their labels follow accordingly: in Eqs.~\eqref{eq:skewness_label} $Q$, $PB$ ans $LSS$ respectively stand for {\it Quadratic}, {\it Post-Born} and {\it Large-Scale-Structure}  since
\begin{itemize}
\item{$\mu^Q_3$ takes into account the non-Gaussianity coming from the quadratic term $\sigma^{(1)\,2}/2$ in Eq.~\eqref{eq:52},}
\item{$\mu^{PB}_3$ contains all the relevant terms due to the Post-Born corrections to the $d_L(z)$, such as multi-lens effects,}
\item{$\mu^{LSS}_3$ catches the non-Gaussianities arising from the bispectrum of $\overline{\delta^{(2)}\delta^{(1)}\delta^{(1)}}$.}
\end{itemize}

Understanding how much the finite bin effect may mimic non-Gaussian behavior beyond the fundamental non-linearities is of interest {\it per se}. We will present this study in a forthcoming work. For the rest of this work, we will explicitly compute and numerically evaluate the amplitude of the expected skewness in the infinitesimal redshift bin case, where the only relevant effect is the one due to intrinsic non-Gaussianities in the inhomogeneities.

\section{Analytic expressions}
\label{sec:analytic}
In this section, we provide the explicit expression for the leading order terms of the skewness in the infinitesimal redshift bin limit. First let us remark that our derivation so far is completely geometrical without any assumption on the underlying theory of gravity, with the only assumption that light propagates along null geodesics. At this point we need to use General Relativity to relate metric and matter perturbations.
We first provide some general preliminaries needed to follow our derivations. Technical details are reported in the Appendix~\ref{app:analytic}. In order to compute the different contribution to the third moment defined in Eqs.~\eqref{eq:skewness_label}, it is convenient to introduce the generalized Hankel transform of the matter power spectrum at two given times $\eta_1$ and $\eta_2$, namely  $P(k,\eta_1,\eta_2)\equiv P(k)D_1(\eta_1)D_1(\eta_2)$, as
\beq
J^n_\ell \left( \eta_1,\eta_2 \right) = \int \frac{d k }{2 \pi^2} k^2 P \left( k,\eta_1,\eta_2 \right) \frac{j_\ell \left( k |r_1-r_2| \right) }{\left( k |r_1-r_2| \right)^n}\,,
\label{eq:Hankel}
\eeq
where $r_i\equiv r(\eta_i)=\eta_0-\eta_i$, $j_\ell(x)$ is the $\ell$-th order spherical Bessel function and $D_1$ the growth function normalized to unity today. The reason why we decide to adopt the definition \eqref{eq:Hankel} for the Hankel transform is that it can be easily generalised to the science case of non-linear matter power spectrum, as we will discuss in Sect.~\ref{sec:numerics}.

In evaluating Eqs.~\eqref{eq:skewness_label}, we encounter the following terms
\bea
\overline{\Delta_2 \psi \left( r_1 ,\bn \right) \Delta_2 \psi \left( r_2, \bn \right)}
&=&\frac{9\,r_1 r_2\,\HH_0^4\,\Omega_{m0}^2}{a(r_1) a(r_2)}
\left[2 r_1 r_2 J^2_2 \left(\eta_1,\eta_2 \right)
\right.\nonumber\\
&&\left.+J^3_1 \left(\eta_1,\eta_2\right)  (r_1-r_2)^2\right]
\equiv\mathcal{L}\left( r_1,r_2 \right)\,,
\nonumber\\
\bar \gamma_0^{a b}\overline{ \partial_b \Delta_2 \psi \left( r_1 ,\bn \right) \partial_a \psi \left( r_2, \bn \right)}
&=&-\mathcal{L}\left( r_1,r_2 \right)\,,
\nonumber\\
\overline{\Delta_2 \left(\bar \gamma_0^{a b} \partial_a \psi \left( r_1 ,\bn \right) \partial_b \psi \left( r_2 ,\bn \right) \right)  } &=& 0\,,
\nonumber\\
\overline{\Delta_2\psi\left( r_1,\bn \right)\pa_a\psi\left( r_2,\bn \right)}
&=&\overline{\psi\left( r_1,\bn \right)\pa_a\Delta_2\psi\left( r_2,\bn \right)}=0\,.
\label{eq:2point}
\eea
We remark that 
\beq
\overline{\Delta_2 \psi \left( r_1 ,\bn \right) \Delta_2 \psi \left( r_2, \bn \right)}  + \bar \gamma_0^{a b}\overline{ \partial_b \Delta_2 \psi \left( r_1 ,\bn \right) \partial_a \psi \left( r_2, \bn \right)} =0\,,
\eeq
whereas second last of Eqs.~\eqref{eq:2point} vanishes since
\beq
\overline{\Delta_2 \left(\bar \gamma_0^{a b} \partial_a \psi \left( r_1 ,\bn \right) \partial_b \psi \left( r_2 ,\bn \right) \right)  } = \Delta_2 \overline{\left(\bar \gamma_0^{a b}  \partial_a \psi \left( r_1 ,\bn \right) \partial_b \psi \left( r_2 ,\bn \right) \right)  }
\eeq
and $\overline{\left(\bar \gamma_0^{a b}  \partial_a \psi \left( r_1 ,\bn \right) \partial_b \psi \left( r_2 ,\bn \right) \right)  }$ can not depend on the direction $\bn$ due to statistical isotropy.
For the same reason
\beq
\overline{\Delta_2 \psi^{(2)}} \propto \overline{\Delta_2 \psi^{2}} = \Delta_2 \overline{\psi^2} =0\,.
\label{eq:67}
\eeq
Last equality in Eq.~\eqref{eq:2point} can be understood in the following way: they always involve an odd number of angular derivatives and hence they vanish due to statistical isotropy, since they naturally introduce a preferred direction, which returns to 0 when the ensemble average acts.

The 2-point correlation functions in Eqs.~\eqref{eq:2point} are enough also to evaluate the 4-point correlation functions of interest for us. With the aim of Wick theorem, indeed, we can write
\bea
&&\overline{\Delta_2\psi(r_1,{\bf n})\Delta_2\psi(r_2,{\bf n})
\Delta_2 \psi \left( r_3 ,\bn \right) \Delta_2 \psi \left( r_4 ,\bn \right)}
\nonumber\\
&=&\mathcal{L}\left( r_1,r_2 \right)\mathcal{L}\left( r_3,r_4 \right)
+\mathcal{L}\left( r_1,r_3 \right)\mathcal{L}\left( r_2,r_4 \right)
+\mathcal{L}\left( r_1,r_4 \right)\mathcal{L}\left( r_3,r_2 \right)\,.
\label{eq:4point1}
\eea
For the sake of completeness, we report in Appendix~\ref{app:analytic} the derivation of Eq.~\eqref{eq:4point1}. In a similar manner, we obtain also that
\bea
&&\overline{\Delta_2\psi(r_1,{\bf n})\Delta_2\psi(r_2,{\bf n})
\Delta_2 \left(\bar \gamma_0^{a b} \partial_a \psi \left( r_3 ,\bn \right) \partial_b \psi \left( r_4 ,\bn \right) \right) }
\nonumber\\
&=&\mathcal{L}\left( r_1,r_3 \right)\mathcal{L}\left( r_2,r_4 \right)
+\mathcal{L}\left( r_1,r_4 \right)\mathcal{L}\left( r_2,r_3 \right)\,.
\label{eq:4point2}
\eea
It is interesting to notice that only two permutations survive in Eq.~\eqref{eq:4point2} as a consequence of the last two equalities in Eqs.~\eqref{eq:2point}. Looking at the structure of Eq.~\eqref{eq:4point2}, we notice that only the permutations mixing $\Delta_2\psi$ with one of the term of $\Delta_2(\pa\psi)^2$ survive. Finally, last 4-point correlation function of our interest is
\bea
&&\overline{\Delta_2\psi(r_1,{\bf n})\Delta_2\psi(r_2,{\bf n})
\bar \gamma_0^{a b}\partial_b \Delta_2 \psi \left( r_3 ,\bn \right) \partial_a \psi \left( r_4, \bn \right)}
\nonumber\\
&=&
\overline{\Delta_2\psi(r_1,{\bf n})\Delta_2\psi(r_2,{\bf n})}
\,\overline{\bar \gamma_0^{a b}\partial_b \Delta_2 \psi \left( r_3 ,\bn \right) \partial_a \psi \left( r_4, \bn \right)}
=-\mathcal{L}\left( r_1,r_2 \right)\mathcal{L}\left( r_3,r_4 \right)\,.
\label{eq:4point3}
\eea
The interesting thing about Eq.~\eqref{eq:4point3} is that only one permutation survives in the final result and this is again a consequence of last of Eqs.~\eqref{eq:2point}. The structure of Eq.~\eqref{eq:4point3} also exhibits a complete factorization of the 2-point correlation function in the form $\left( \Delta_2\psi \right)^2$. This situation is opposite to what we have shown for Eq.~\eqref{eq:4point2}, where only mixed terms survive in the final permutations. The impact of these differences in the ultimate evaluation of $\mu_3$ will be to significantly reduce the final number of non-null terms for the skewness, as we will show in Sect.~\ref{sec:PB}.

These analytic preliminaries are enough to provide the explicit expressions for $\mu^Q_3$, $\mu^{PB}_3$ in the infinitesimal bin case. The evaluation of $\mu^{LSS}_3$ is more delicate and will be treated in a specific way.

\subsection{$\mu^Q_3$: quadratic terms}
\label{sec:quad}
The starting point for the computation of $\mu^Q_3$ is its expression in Eqs.~\eqref{eq:skewness_label}. To our hand, we first evaluate the variance in the small bin limit. By making use of Eqs.~\eqref{eq:variance}, \eqref{eq:52}, and \eqref{eq:2point}, we get that
\beq
\sigma^2=\int_0^{r_s}dr_1\frac{r_1-r_s}{r_1r_s}
\int_0^{r_s}dr_2\frac{r_2-r_s}{r_2r_s}\mathcal{L}(r_1,r_2)\,.
\label{eq:var}
\eeq
In the same way, thanks to Eq.~\eqref{eq:4point1}, we have that
\bea
\overline{I[\sigma^{(1)\,4}]}&=&
\int_0^{r_s}dr_1\frac{r_1-r_s}{r_1r_s}
\int_0^{r_s}dr_2\frac{r_2-r_s}{r_2r_s}
\int_0^{r_s}dr_3\frac{r_3-r_s}{r_3r_s}
\int_0^{r_s}dr_4\frac{r_4-r_s}{r_4r_s}
\nonumber\\
&&\times
\left[\mathcal{L}\left( r_1,r_2 \right)\mathcal{L}\left( r_3,r_4 \right)
+\mathcal{L}\left( r_1,r_3 \right)\mathcal{L}\left( r_2,r_4 \right)
+\mathcal{L}\left( r_1,r_4 \right)\mathcal{L}\left( r_3,r_2 \right)\right]
\nonumber\\
&=&3\,\left(\sigma^2\right)^2\,.
\label{eq:four}
\eea
Hence, the combination of Eqs.~\eqref{eq:52}, \eqref{eq:var} and \eqref{eq:four} we get the quite simple result
\beq
\mu^Q_3=3\left( \sigma^2 \right)^2\,.
\label{eq:quadratic_skewness}
\eeq

It is worth noticing what happens for the leading term of the standardized third moment. For the quadratic terms coming from Eq.~\eqref{eq:quadratic_skewness}, we get that
\beq
\kappa^Q_3
\equiv\frac{\mu^Q_3}{\left(\sigma^2\right)^{3/2}}
=3\,\sigma\,,
\label{eq:k3Q}
\eeq
namely the skewness of the distance-redshift relation due to the quadratic corrections is proportional to the dispersion of $d_L(z)$. This result already allows us to estimate the amplitude of $\kappa^Q_3$. Indeed, in \cite{Ben-Dayan:2013nkf} the dispersion for the distance-redshift relation has been estimated for the lensing contribution at higher redshift to be $\sim 1\%$. This evaluation takes into account the non-linear power spectrum for the gravitational potential as provided by the Halofit model~\cite{Takahashi:2012em}.

\subsection{$\mu^{PB}_3$: Post-Born corrections}
\label{sec:PB}
For the evaluation of the Post-Born terms $\mu^{PB}_3$ in Eq.~\eqref{eq:skewness_label}, we first look at the expression of $\Sigma^{(2)}$ in Eq.~\eqref{eq:53} and treat its terms separately. We divide $\Sigma^{(2)}=\Sigma^{(2)}_{sep}+\Sigma^{(2)}_{mix}$, where we defined
\bea
\Sigma^{(2)}_{sep}&\equiv&2\int_0^{r_s}\,dr\frac{r-r_s}{rr_s}\pa_b\left[ \Delta_2\psi(r) \right]
\int_0^{r_s}dr\,\frac{r-r_s}{rr_s}\bar\gamma_0^{ab}\pa_a\psi(r)\nonumber\\
&&+2\,\int_0^{r_s}dr\left\{ \gamma_0^{ab}\pa_b\left[ \int_0^rdr'\,\psi(r') \right]
\int_0^rdr'\,\frac{r'-r}{rr'}\pa_a\Delta_2\psi(r') \right\}\,,\nonumber\\
\Sigma^{(2)}_{mix}&\equiv&\int_0^{r_s}dr\,\frac{r-r_s}{rr_s}\Delta_2\left[ \gamma_0^{ab}\pa_a\left( \int_0^rdr'\,\psi(r') \right)\pa_b\left( \int_0^rdr'\,\psi(r') \right) \right]\,,
\eea
such the total contribution to $\mu^{PB}_3$ is given by
\beq
\mu^{PB}_{3\,sep}=3\left\{\overline{I[\sigma^{(1)\,2} \Sigma^{(2)}_{sep}]}
-\sigma^2\overline{I[\Sigma^{(2)}_{sep}]}\right\}\,,
\label{eq:skewness_sep}
\eeq
and
\beq
\mu^{PB}_{3\,mix}=3\left\{\overline{I[\sigma^{(1)\,2} \Sigma^{(2)}_{mix}]}
-\sigma^2\overline{I[\Sigma^{(2)}_{mix}]}\right\}\,.
\label{eq:skewness_mix}
\eeq

Thanks to Eqs.~\eqref{eq:4point2} and \eqref{eq:4point3}, $\mu^{PB}_{3\,sep}$ and $\mu^{PB}_{3\,mix}$ can be readily evaluated. In fact, since the factorization in Eq.~\eqref{eq:4point3} does not mix any term of $\Sigma^{(2)}_{sep}$ with $\sigma^{(1)}$, we can immediately factorize $\overline{I\left[\sigma^{(1)\,2} \Sigma^{(2)}_{sep}\right]}=\overline{I\left[\sigma^{(1)\,2}\right]}\overline{I\left[\Sigma^{(2)}_{sep}\right]}$ as well. Then, using Eq.~\eqref{eq:variance}, this automatically returns $\mu^{PB}_{3\,sep}=0$. For what concerns $\mu^{PB}_{3\,mix}$, we first notice that $\overline{I\left[\Sigma^{(2)}_{mix}\right]}=0$ as a consequence of Eqs.~\eqref{eq:2point}. Moreover, from the structure of the 4-point correlation function already discussed after Eq.~\eqref{eq:4point2}, we obtain
\bea
\mu^{PB}_{3\,mix}
=6\int_0^{r_s} dr_1 \frac{r_1-r_s}{r_1 r_s}
\int_0^{r_s} dr_2 \frac{r_2-r_s}{r_2 r_s}
\int_0^{r_s} \frac{dr}{r^2} \frac{r-r_s}{r r_s}\int_0^r dr_3\int_0^r dr_4
\,\mathcal{L}\left( r_1,r_3 \right)\mathcal{L}\left( r_2,r_4 \right)\,.
\nonumber\\
\eea
Hence the total contribution of the Post-Born corrections to the third moment is
\beq
\mu^{PB}_3 = \mu^{PB}_{3\,mix}\,,
\eeq
and the skewness is
\beq
\kappa^{PB}_3\equiv\frac{\mu^{PB}_3}{\left( \sigma^2 \right)^{3/2}}
=\frac{\mu^{PB}_{3\,mix}}{\left( \sigma^2 \right)^{3/2}}\,.
\label{eq:k3PB}
\eeq

The structure of $\mu^{PB}_3$ reveals a sequence of five nested line-of-sight integrals, consistent with our expectations from Post-Born corrections. These corrections effectively incorporate the non-linearities arising from multi-lens effects. 
To address numerical challenges and minimize the number of integrals, we propose an approximation based on the observation that $\mathcal{L}(r_1,r_2)$ exhibits a strong peak near $r_1 \approx r_2$. Thus, we can approximate $\mathcal{L}(r_1,r_3)$ as $\sim \delta_{D}(r_1 - r_3)$, simplifying the calculations. We then
have
\begin{equation}
\int_0^r dr_3\mathcal{L}(r_1,r_3)\approx
f(r_1)\Theta(r-r_1)\,,
\label{eq:PB_approximation}
\end{equation}
where $f(r_1)$ is a function to be determined in a numerical way and $\Theta(x)$ is the Heaviside step function.
This means that
\begin{equation}
\mu^{PB}_{3\,mix}
\approx 6\int_0^{r_s} dr_1 \frac{r_1-r_s}{r_1 r_s}
\int_0^{r_s} dr_2 \frac{r_2-r_s}{r_2 r_s}
\int_0^{r_s} \frac{dr}{r^2} \frac{r-r_s}{r r_s}
f(r_1)\Theta(r-r_1)
f(r_2)\Theta(r-r_2)
\,.
\end{equation}
For what concerns $f(r_1)$, since the latter is not a function of $r$ and the integrand $\mathcal{L}$ contributes to the integral in Eq.~\eqref{eq:PB_approximation} only around $r_3\approx r_1$, we have that integral \eqref{eq:PB_approximation} is independent of the value of $r$ as long as $r>r_1$. To this end, then, we can evaluate the function $f$ once for all by choosing an $r$ equal or larger than a given comoving distance $r^*$ well-beyond the highest redshift that we investigate. For our purposes, we have chosen $r^*=r(z=4)$ but the final results are clearly independent of this choice. In this way, we have that
\begin{equation}
f(r_1)=\int_0^{r^*} dr_3\mathcal{L}(r_1,r_3)\,.
\end{equation}
Numerical results within this approximation will be discussed in Sect.~\ref{sec:numerics}.

\subsection{$\mu^{LSS}_3$: the role of the bispectrum}
The evaluation of $\mu^{LSS}_3$ is the more demanding, since involves the 3-point function
\begin{equation} \label{eq:3point_Psi}
\overline{\Delta_2 \psi \left( r_1 ,\bn \right) \Delta_2 \psi \left( r_2, \bn \right) \Delta_2 \Psi^{(2)} \left( r_3, \bn \right)}\, ,
\end{equation}
where we have defined $\Psi^{(2)}\equiv\frac{1}{2}\left(\psi^{(2)}+\phi^{(2)}\right)$.
To evaluate this correlation, we use again the Poisson equation to relate $\Psi^{(2)}$ to the second-order matter fluctuations $\delta^{(2)}$.
We then have that 
\bea
&&
\overline{\Delta_2 \psi \left( r_1 ,\bn \right) \Delta_2 \psi \left( r_2, \bn \right) \Delta_2 \Psi^{(2)} \left( r_3, \bn \right)}
\nonumber\\
&=& \int  \frac{d^3k_1\,d^3k_2\,d^3k_3}{\left( 2 \pi \right)^{9}}
\overline{\psi(k_1,\bn) \psi(k_2,\bn) \Psi^{(2)} (k_3,\bn)}
\Delta_2 e^{i \bk_1 \cdot \bn r_1} 
\Delta_2 e^{i \bk_2 \cdot \bn r_2}  
\Delta_2 e^{i \bk_3 \cdot \bn r_3}
\nonumber\\
&=&-\,C \int  \frac{d^3k_1\,d^3k_2\,d^3k_3}{\left( 2 \pi \right)^{6}}  
\delta_D \left( \bk_1 + \bk_2 + \bk_3 \right)
\frac{B\left( k_1 , k_2,k_3,r_1,r_2,r_3 \right)}{k_1^2 k_2^2 k_3^2 }
 \nonumber\\
&&\times
\Delta_2 e^{i \bk_1 \cdot \bn r_1} 
\Delta_2 e^{i \bk_2 \cdot \bn r_2}  
\Delta_2 e^{i \bk_3 \cdot \bn r_3}\,,
\label{eq:3point}
\eea
where\footnote{This factor follows from the fact that the transfer function for $\delta^{(2)}$ is linked to the transfer function of $\Psi^{(2)}$ by the Poisson equation, which preserves the form as the linear one reported in Eq.~\eqref{eq:transf_dict}.}
$C\equiv\frac{27}{8}\frac{\HH_0^6\,\Omega_{m0}^3}{a(r_1)a(r_2)a(r_3)}$, and we have introduced the (non-symmetrized) bispectrum
\begin{equation}
    \overline{\delta(\bk_1,r_1) \delta(\bk_2,r_2) \delta^{(2)}(\bk_3,r_3)} = \left( 2 \pi \right)^3 \delta_D \left( \bk_1 + \bk_2 + \bk_3 \right) B\left( k_1, k_2, k_3, r_1, r_2, r_3 \right)\,,
\end{equation}
which at tree-level is given by
\begin{equation}
    B\left( k_1, k_2, k_3, r_1, r_2, r_3 \right) = 2 D_1\left( r_1 \right) D_1 \left( r_2 \right) D_1^2 \left( r_3 \right) F_2 \left( k_1, k_2, k_3 \right) P\left( k_1 \right) P \left( k_2 \right) \, ,
\end{equation}
with
\begin{equation}
    F_2 \left( k_1, k_2 ,k_3 \right) = \frac{5}{7} +\frac{1}{4} \frac{k_3^2-k_1^2-k_2^2}{k_1 k_2} \left( \frac{k_1}{k_2} +\frac{k_2}{k_1} \right)+ \frac{1}{14} \left( \frac{k_3^2-k_1^2-k_2^2}{k_1 k_2} \right)^2 \, .
\end{equation}
Finally, accounting for the two other permutations of Eq.~\eqref{eq:3point_Psi} we need to simply replace $B (k_1, k_2, k_3 )$ with its symmetrized version
\begin{eqnarray}
    B_{\rm sym} (k_1,k_2, k_3 ,r_1,r_2,r_3) &=& B (k_1, k_2, k_3,r_1, r_2,r_3 ) +B (k_2, k_3, k_1, r_2, r_3, r_1 )
    \nonumber \\
    && +B (k_3, k_1, k_2 , r_3, r_1, r_2 ) \, .
\end{eqnarray}
From Eq.~\eqref{eq:3point}, we can write the 3-point function as 
\bea
&& \hspace{-0.5cm}\overline{\Delta_2 \psi \left( r_1 ,\bn \right) \Delta_2 \psi \left( r_2, \bn \right) \Delta_2 \Psi^{(2)} \left( r_3, \bn \right)} +\circlearrowleft
\nonumber \\
&=& C(r_1,r_2,r_3)\frac{8 \left( 4 \pi \right)}{\left(2\pi \right)^6}  \sum_{\ell_1\ell_2\ell_3} \ell_1 \left( \ell_1 +1 \right) \ell_2 \left( \ell_2 +1 \right) \ell_3 \left( \ell_3 +1 \right)\nonumber\\
&&\times\left( \begin{array}{ccc}
\ell_1 & \ell_2 & \ell_3 \\ 0 & 0 & 0 
\end{array} \right)^2 
\left( 2\ell_1 +1 \right)\left( 2\ell_2 +1 \right)\left( 2\ell_3 +1 \right)
\int  dk_1\,dk_2\,dk_3\,dx\,x^2
B_{\rm sym}\left( k_1 , k_2,k_3,r_1,r_2,r_3 \right)
\nonumber\\
&&\times 
j_{\ell_1 }\left( k_1 r_1 \right)j_{\ell_1} \left( k_1 x \right)
j_{\ell_2 }\left( k_2 r_2 \right)j_{\ell_2} \left( k_2 x \right)
j_{\ell_3 }\left( k_3 r_3 \right)j_{\ell_3} \left( k_3 x \right)\,,
\label{eq:bis1}
\eea
where
\beq
\left( \begin{array}{ccc}
\ell_1 & \ell_2 & \ell_3 \\ m_1 & m_2 & m_3 
\end{array} \right)\,,
\eeq
denotes the so-called 3-j Wigner symbol, which is non-null only when $m_1+m_2+m_3=0$ and $(\ell_1,\ell_2,\ell_3)$ satisfy the triangular inequality. The presence of these symbols selects only specific shapes for the sums in $\ell$-space. The detailed derivation of Eq.~\eqref{eq:bis1} can be found in Appendix~\ref{app:analytic}. Here we directly report the result which are useful for us to discuss the underlying physics. Despite its compact form, Eq.~\eqref{eq:bis1} is still quite demanding for our scopes. Indeed, we still need to evaluate the contribution to the third moment as given in Eq.~\eqref{eq:skewness_label}. This implies that three line-of-sight integrals must be performed over Eq.~\eqref{eq:bis1}. In total, then, we have to numerically compute seven integrals for each non-null set of $(\ell_1,\ell_2,\ell_3)$ and this is quite unpractical from the numerical viewpoint.

In order to face the dramatic need for reducing the number of integrals, we invoke the Limber approximation for each integral in $k$-space in Eq.~\eqref{eq:bis1}. Thanks to it, in fact, we can make the following approximation~\cite{1953ApJ...117..134L,Limber:1954zz,LoVerde:2008re}
\bea
\frac{2}{\pi}\int dk k^2 g(k)j_\ell(k r)j_\ell(k s)
\approx \frac{\delta_D\left( r-s \right)}{r^2}g\left( \frac{\ell+1/2}{r} \right)\,,
\label{eq:Limber}
\eea
whenever the function $g$ does not vary too much rapidly. In this way, the approximation \eqref{eq:Limber} can be used three times to get rid of the $k$-space integrals. This reduces the number of remaining integrations to four. Hence, Eq.~\eqref{eq:bis1} becomes 
\bea
&&\overline{\Delta_2 \psi \left( r_1 ,\bn \right) \Delta_2 \psi \left( r_2, \bn \right) \Delta_2 \Psi^{(2)} \left( r_3, \bn \right)} +\circlearrowleft
\approx 
\nonumber \\
&\approx&
C\frac{8 \left( 4 \pi \right)}{\left(2\pi\right)^6} \frac{\pi^3}{2^3}  \sum_{\ell_1\ell_2\ell_3} \ell_1 \left( \ell_1 +1 \right) \ell_2 \left( \ell_2 +1 \right) \ell_3 \left( \ell_3 +1 \right)\nonumber\\
&&\times\left( \begin{array}{ccc}
\ell_1 & \ell_2 & \ell_3 \\ 0 & 0 & 0 
\end{array} \right)^2 
\left( 2\ell_1 +1 \right)\left( 2\ell_2 +1 \right)\left( 2\ell_3 +1 \right)
\int \frac{dx\,x^2}{r^2_1r^2_2r^2_3}
\delta_D\left( x-r_1 \right)\delta_D\left( x-r_2 \right)\delta_D\left( x-r_3 \right)
\nonumber\\
&&\times \frac{B_{\rm sym}\left( \widetilde{k_1},\widetilde{k_2},\widetilde{k_3} ,r_1,r_2,r_3\right)}{\widetilde{k_1}^2\widetilde{k_2}^2\widetilde{k_3}^2}
\,,
\label{eq:bis2}
\eea
where $\widetilde{k_i}\equiv \frac{\ell_i+1/2}{r_i}$. The three delta Dirac distributions in Eq.~\eqref{eq:bis2} simplify the three line-of-sight integrals in the last relation of Eq.~\eqref{eq:skewness_label}, leading to the final expression for the LSS contribution to the third moment
\bea
\mu^{LSS}_3&=&
\frac{1}{\left(4\pi\right)^2}\sum_{\ell_1\ell_2\ell_3} \ell_1 \left( \ell_1 +1 \right) \ell_2 \left( \ell_2 +1 \right) \ell_3 \left( \ell_3 +1 \right)\left( \begin{array}{ccc}
\ell_1 & \ell_2 & \ell_3 \\ 0 & 0 & 0 
\end{array} \right)^2\left( 2\ell_1 +1 \right)\left( 2\ell_2 +1 \right)\left( 2\ell_3 +1 \right)\nonumber\\
&&\times\int_0^{r_s}dx\, \frac{C(x,x,x)}{x^4}
\left(\frac{x-r_s}{x\,r_s}\right)^3
\frac{B_{\rm sym}\left( \widetilde{k_1},\widetilde{k_2},\widetilde{k_3},x,x,x \right)}{\widetilde{k_1}^2\widetilde{k_2}^2\widetilde{k_3}^2}
\nonumber \\
&=& 
\frac{4}{ \pi^2}
\sum_{\ell_1\ell_2\ell_3} 
\frac{\ell_1 \left( \ell_1 +1 \right)}{2 \ell_1 +1}
\frac{\ell_2 \left( \ell_2 +1 \right)}{2 \ell_2 +1}
\frac{\ell_3 \left( \ell_3 +1 \right)}{2 \ell_3 +1}
\left( \begin{array}{ccc}
\ell_1 & \ell_2 & \ell_3 \\ 0 & 0 & 0 
\end{array} \right)^2
\nonumber\\
&&\times\int_0^{r_s}dx\, \frac{C(x,x,x)}{x}
\left(\frac{x-r_s}{r_s}\right)^3
B_{\rm sym}\left( \widetilde{k_1},\widetilde{k_2},\widetilde{k_3},x,x,x \right) 
\,,
\label{eq:bis3}
\eea
where we used Eqs.~\eqref{eq:52} and \eqref{eq:53}, and we defined now $\widetilde{k_i}= \frac{\ell_i+1/2}{x}$ and $\eta_x\equiv\eta_0-x$. Then, we have that the skewness due to the non-linear structures is
\beq
\kappa^{LSS}_3=\frac{\mu^{LSS}_3}{\left(\sigma^2\right)^{3/2}}\,.
\label{eq:k3LSS}
\eeq

\section{Numerical results}
\label{sec:numerics}
In this section, we discuss the numerical integrations of the analytical results for the skewness given by Eqs.~\eqref{eq:k3Q}, \eqref{eq:k3PB} and \eqref{eq:k3LSS}. We work with the fiducial cosmology given by the $\Lambda$CDM parameters $\Omega_{c}=0.2638$, $\Omega_b=0.04827$ $h=0.67556$, $A_s=2.215 \times 10^{-9}$ and $n_s=0.9619$ in accordance with \cite{Adamek:2018rru}.

\subsection{The smoothing scale}
The evaluation of the skewness, or any other 1-point function, is strongly sensitive to the non-linear nature of structure formation. However, by working in perturbation theory we know that our prediction will fail beyond the mildly non-linear scale. Therefore, our prediction can be compared to real observations only when the non-linear effects are filtered out, by introducing a smearing scale in real space of size $\rho$. In this regard, we use a spherical top-hat window function $\mathcal{W}(r,\rho)$ in real space. This procedure introduces a window function $W(k,\rho)$ in Fourier space given by
\begin{equation}
W(k,\rho)= 3\,\frac{j_1\left( k\rho \right)}{k\rho}\,,
\end{equation}
which affects the power spectrum as
\begin{equation}
P(k,\eta_1,\eta_2)\rightarrow P_\rho(k,\eta_1,\eta_2)
\equiv P(k,\eta_1,\eta_2) W^2(k,\rho)\,.
\label{eq:windowed_PS}
\end{equation}
With such smoothing scale, all the integrals over the momentum $k$ converge quickly and the choice of $k_{\rm max}$ become numerically irrelevant.

For the term $\kappa_3^{LSS}$ we will consider also some non-linear fitting expression of the bispectrum, in particular BiHalofit~\cite{Takahashi:2019hth}. For this reason, to smooth out small scales non-linearities we apply directly the window functions as follows
\bea
&&
\int d^3r'_1 d^3r'_2 d^3r'_3
\mathcal{W}\left(\left| \br_1 - \br'_1 \right|, \rho \right)
\mathcal{W}\left(\left| \br_2 - \br'_2 \right|, \rho \right) 
\mathcal{W}\left(\left| \br_3 - \br'_3 \right|, \rho \right)
\nonumber\\
&&\times
\overline{\Delta_2 \psi \left( r'_1 ,\bn \right) \Delta_2 \psi \left( r'_2, \bn \right) \Delta_2 \Psi^{(2)} \left( r'_3, \bn \right)}
\nonumber\\
&=& \int  \frac{d^3k_1\,d^3k_2\,d^3k_3}{\left( 2 \pi \right)^{9}} W(k_1, \rho)W(k_2, \rho) W(k_3, \rho)
\overline{\psi(k_1,\bn) \psi(k_2,\bn) \Psi^{(2)} (k_3,\bn)}
\nonumber\\
&&\times
\Delta_2 e^{i \bk_1 \cdot \bn r_1} 
\Delta_2 e^{i \bk_2 \cdot \bn r_2}  
\Delta_2 e^{i \bk_3 \cdot \bn r_3}\,.
\eea
To include the window function we simply need to replace eq.~\eqref{eq:bis3} with
\bea
\mu^{LSS}_3&=&
\frac{4}{ \pi^2}
\sum_{\ell_1\ell_2\ell_3} 
\frac{\ell_1 \left( \ell_1 +1 \right)}{2 \ell_1 +1}
\frac{\ell_2 \left( \ell_2 +1 \right)}{2 \ell_2 +1}
\frac{\ell_3 \left( \ell_3 +1 \right)}{2 \ell_3 +1}
\left( \begin{array}{ccc}
\ell_1 & \ell_2 & \ell_3 \\ 0 & 0 & 0 
\end{array} \right)^2\\
&&\times\int_0^{r_s}dx\, \frac{C(x,x,x)}{x}
\left(\frac{x-r_s}{r_s}\right)^3
B_{\rm sym}\left( \widetilde{k_1},\widetilde{k_2},\widetilde{k_3},x,x,x \right) {  W(\tilde k_1, \rho)W(\tilde k_2, \rho) W(\tilde k_3, \rho)}
\,,\nonumber
\eea
In the following, we will investigate how suitable choices for $\rho$ can be made: we will consider the cases of $\rho$ equal to 5, 10 and 20 Mpc/$h$.

\subsection{Quadratic and Post-Born terms}
\begin{figure}[ht!]
\centering
\includegraphics[scale=0.31]{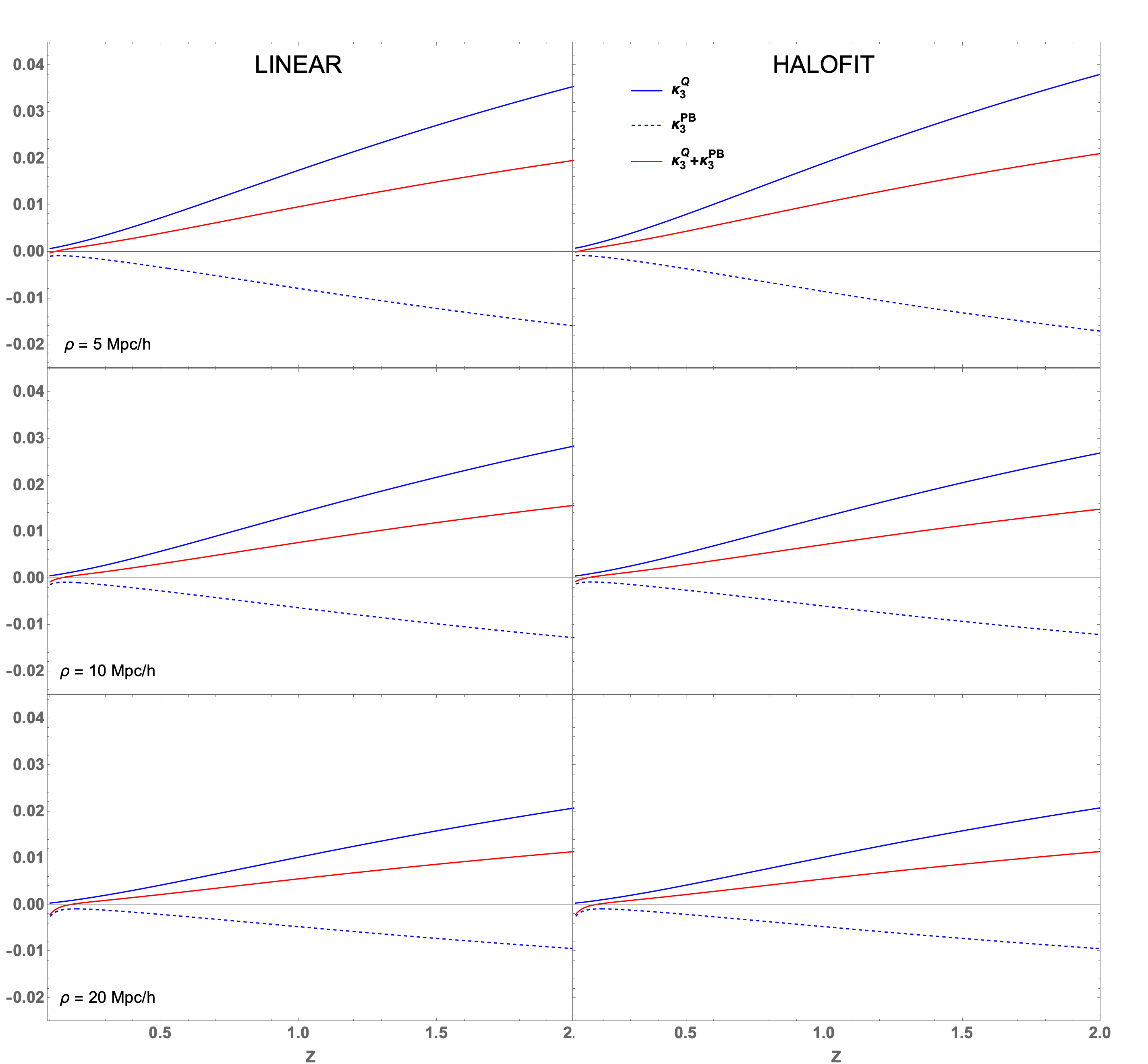}
\caption{Quadratic (blue), Post-Born (blue dashed) terms and their sum (red) for the skewness, evaluated at with a coarse-graining scale $\rho=5$ Mpc/$h$ (top), $\rho=10$ Mpc/$h$ (center) and $\rho=20$ Mpc/$h$ (bottom). Left panels are obtained with linear power spectrum, whereas right panels consider Halofit model for the matter power spectrum.}
\label{fig:lin}
\end{figure}

We start our numerical investigation by looking at the contributions to the skewness arising from the linear gravitational potential, derived in Sects.~\ref{sec:quad} and \ref{sec:PB}. For this discussion, we will use the cutoff in real space on small scales as prescribed by Eq.~\eqref{eq:windowed_PS}. Results are shown in Fig.~\ref{fig:lin} respectively for the choices of $\rho = 5,\,10,\,20$ Mpc/$h$. For each case, we consider linear matter power spectrum (left panels) and Halofit model for the matter power spectrum (right panels).

With these results, we can spot common features among the different cases. First of all, we notice that quadratic terms in $\kappa^Q_3$ (blue solid lines in Fig.~\ref{fig:lin}) are always positive and lead to an increasing skewness with redshift. The order of magnitude of this term reaches about $10^{-2}$. In this regard, the positive sign was already expected just by looking at the structure of $\kappa^Q_3$ in Eq.~\eqref{eq:k3Q}, since it is just proportional to the dispersion of the $d_L(z)$ distribution. 
We have also checked that our framework is in a good agreement with \cite{Ben-Dayan:2013nkf} when we push the coarse-graining scale up to $0.3 \ {\rm Mpc}/h$, roughly corresponding to the UV cutoff $k_{UV}=10 \ h/{\rm Mpc}$ used in \cite{Ben-Dayan:2013nkf}.

For what concerns the Post-Born contribution to the skewness, namely $\kappa^{PB}_3$ in Eq.~\eqref{eq:k3PB}, blue dashed lines in Fig.~\ref{fig:lin} show two important features. First of all, the Post-Born contribution to the skewness is always negative. This is important to be addressed since $\kappa^{PB}_3$ in Eq.~\eqref{eq:k3PB} does not show any manifest sign, contrary to $\kappa^Q_3$. As a second matter of fact, Post-Born corrections contribute to the skewness with the same order of magnitude of $\kappa^Q_3$. This is somehow in line with the structure of $\kappa^Q_3$ and $\kappa^{PB}_3$, since they just involve different the line-of-sight integrals of the same kernels $\mathcal{L}(r,r')$. Hence, these two features lead to a competitive effect between $\kappa^Q_3$ and $\kappa^{PB}_3$. Indeed, we have a neat effect for the linear gravitational potential skewness, which is still positive but attenuated (solid red lines in Fig.~\ref{fig:lin}). We remark that all the features are shared regardless the value of $\rho$ and whether linear or non-linear matter power spectrum is considered.

\begin{figure}[ht!]
\centering
\includegraphics[scale=0.3]{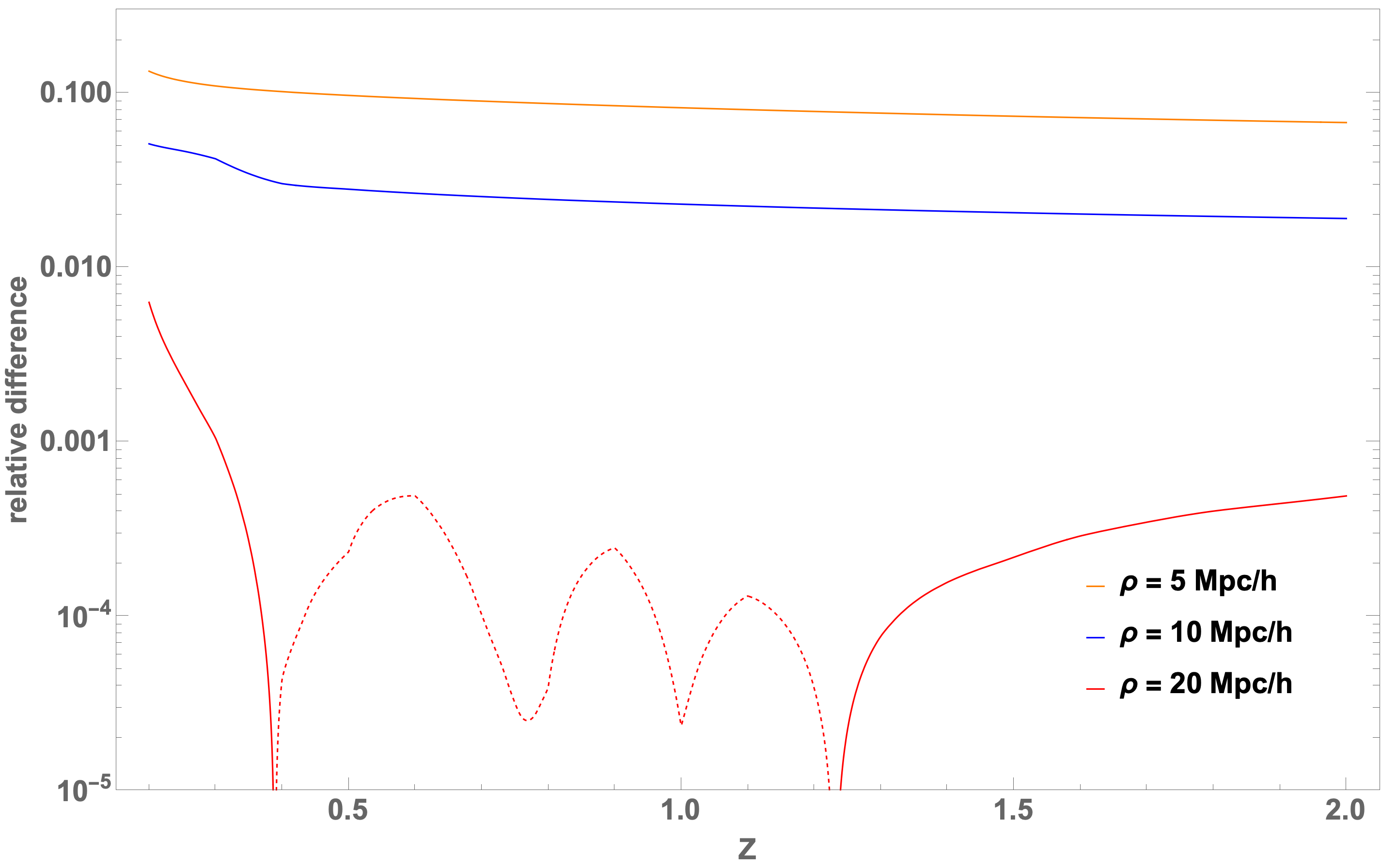}
\caption{Relative difference between the linear and non-linear matter power spectrum for the sum $\kappa_3^Q+\kappa_3^{PB}$. We consider different coarse-graining scales: $\rho=5$ Mpc/$h$ (orange), $\rho=10$ Mpc/$h$ (blue) and $\rho=20$ Mpc/$h$ (red). Dashed curve refers to negative values.}
\label{fig:relative}
\end{figure}

The actual contribution of the non-linear scales to the sum $\kappa^Q_{3}+\kappa^{PB}_{3}$ is quantified in Fig.~\ref{fig:relative}.
Indeed, here we show the relative difference
\beq
\Delta \kappa_3 \equiv 1-\frac{\left(\kappa^Q_3+\kappa^{PB}_3\right)_\text{linear}}{\left(\kappa^Q_3+\kappa^{PB}_3\right)_\text{Halo}}\,,
\eeq
as for different smoothing scales $\rho$. For a value of $\rho=20$ Mpc/$h$, the non-linearities in $P_\rho(k)$ marginally contribute to the total effect with a relative correction of $\sim 0.1\%$. The non-linear scales happen to be more relevant when $\rho$ decreases, namely with a few percent and almost $10\%$ relative correction respectively with a coarse-graining radius of $10$ Mpc/$h$ and $\rho = 5$ Mpc/$h$. The impact of non-linearities in the matter power spectrum also decreases by going to higher redshifts. Let us remark that, in principle, replacing the linear power spectrum in the perturbative expansion with Halofit is not self-consistent. However this can give us a rough estimation of the expected accuracy of our prediction based on perturbation theory for different smoothing scales.

\subsection{Bispectrum}
\label{sec:bispectrum}
The numerical evaluation of the $\kappa^{LSS}_3$ from Eq.~\eqref{eq:k3LSS} requires some subtleties to be accounted for. In fact, from Eq.~\eqref{eq:bis3} we have to specify the value of $\ell_\text{max}$ to effectively compute the contribution of the bispectrum to the skewness.
Since the physical cutoff is determined by the smoothing scale $\rho$, we only need to ensure that $\ell_{\rm max} \gtrsim r_s / \rho$, where the specific value is chosen to guarantee a percent precision.

\begin{figure}[ht!]
\centering
\includegraphics[scale=0.33]{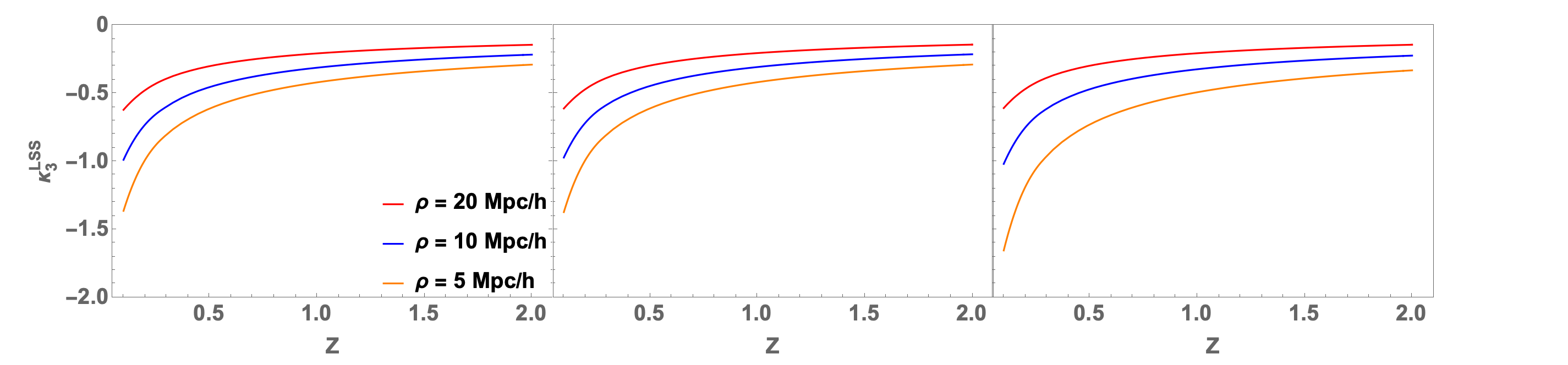}
\caption{Contribution to the skewness given by the bispectrum for the linear (left panel), the Halofit (center panel) and BiHalofit from \cite{Takahashi:2019hth} (right panel) power spectrum for a coarse-graining scale of $\rho=20$ Mpc/$h$ (red), $\rho=10$ Mpc/$h$ (blue) and $\rho=5$ Mpc/$h$ (orange).}
\label{fig:LSS_total}
\end{figure}

We perform numerical investigations by using linear matter power spectrum, Halofit model and the BiHalofit fitting function for the bispectrum provided in \cite{Takahashi:2019hth}. For different values of the coarse-graining scale $\rho$, the numerical results are reported in Fig.~\ref{fig:LSS_total}, where we have adopted linear power spectrum in the left panel, Halofit model in the center and BiHalofit in the right panel.

At first sight, a general feature is that the absolute value of $\kappa^{LSS}_3$ is decreasing with the redshift. In particular, whereas the values at $z\sim 1$ are of order $0.1$, the skewness at lower redshifts varies more and becomes $\mathcal{O}(1)$ when $\rho$ becomes smaller. Moreover, the value is quite insensitive of the kind of spectrum adopted for $\rho = 20$ Mpc/h. This last feature is in line with the fact that the involved scales still evolve almost linearly for this case.

A similar behavior occurs also for the case $\rho = 10$ Mpc/$h$. Even in this case, the value of $\rho$ is still such that non-linear features in the power spectrum are marginally relevant, and then the Halofit and BiHalofit models return results that are almost alongside the ones obtained by the linear power spectrum. However, we notice that the overall amplitude increases quite a lot when we lower the coarse-graining scale from 20 to 10 Mpc/h. As a quantitative instance, the comparison between red and blue curves in Fig.~\ref{fig:LSS_total} at $z=0.1$ exhibits an increasing in the absolute value of the skewness that is of $\sim 40\%$. This confirms that a significant amount of information is encoded in the skewness within the scales of 10 to 20 Mpc/h. However, as the prediction from linear theory does not deviate significantly from the estimations using both Halofit and BiHalofit, we are still within a regime where perturbation theory has not yet failed.

\begin{figure}[ht!]
\centering
\includegraphics[scale=0.3]{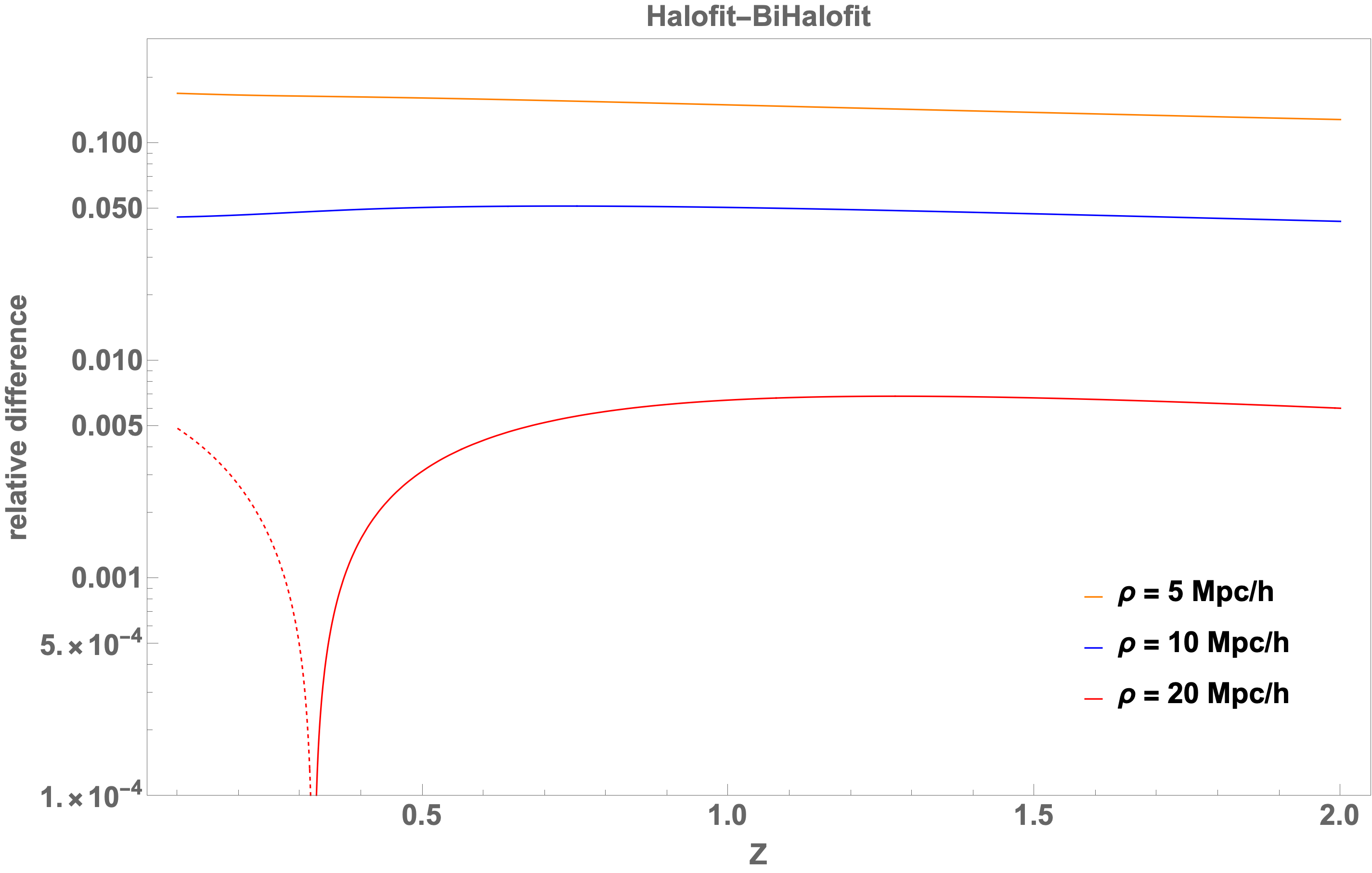}
\caption{Relative difference between the Halofit and BiHalofit matter spectra for the sum $\kappa_3^{LSS}$. We consider different coarse-graining scales: $\rho=5$ Mpc/$h$ (orange), $\rho=10$ Mpc/$h$ (blue) and $\rho=20$ Mpc/$h$ (red). Dashed curve refers to negative values.}
\label{fig:LSS_relative}
\end{figure}

Finally, with orange curves in Fig.~\ref{fig:LSS_total} we report the numerical results for $\rho = 5$ Mpc/$h$. For this case study, we have new emerging features. First of all, we start to appreciate a more prominent difference between the linear power regime and the non-linear one, and this difference is more evident at smaller redshifts. In fact, here we have an enhancement when the BiHalofit model is considered of $\sim 10\%$, as can be appreciated also in Fig.~\ref{fig:LSS_relative}, where we show the analogous of Fig.~\ref{fig:relative} and compare the Halofit fitting formula against the BiHalofit one, but only for $\kappa^{LSS}_3$.
This increasing of $\kappa^{LSS}_3$, as given by Eq.~\eqref{eq:k3LSS}, follows the separate increasing of the two quantities in its ratio, namely $\mu^{LSS}_3$ in Eq.~\eqref{eq:bis3} and $\sigma^3$ as derived from Eq.~\eqref{eq:var}. Given that, we have that the numerator and the denominator of this ratio separately increase when we decrease $\rho$ from 20 to 5 Mpc/$h$.
Hence, by considering the discrepancy between the linear power spectrum and BiHalofit as an indicator of the validity range of perturbation theory, we observe that for $\rho=5$ Mpc/h, the accuracy of perturbation theory cannot be trusted beyond approximately $10\%$.
We also remark that at those redshifts our results are only indicative of the order of magnitude, since other relativistic effects here neglected, such as the Doppler correction might significantly alter the result. This is indeed the case for the dispersion, as shown in Fig.~8 of \cite{Ben-Dayan:2013nkf}. We plan to investigate more in details the contribution of other relativistic effects on small redshift in a forthcoming paper. On the other hand, we also point out that the result is quite stable for $z>0.4$ when we decrease the value of $\rho$.

Two final comments are in order before concluding the discussion devoted to $\kappa^{LSS}_3$. First of all, we recall that our derivation is based on the Limber approximation for the lensing terms. This gave us an appreciable speed-up of the numerical evaluation, since it reduces the number of numerical integrations to one line-of-sight integral, as given in Eq.~\eqref{eq:bis3}. On the other hand, for smaller redshift, the accuracy of this result might be somehow limited, especially because the discrete sum is limited to larger angular scales when the Limber approximation shows some issues. The second comment is about the use of the Halofit model to account for the non-linearities in the matter power spectrum. This is clearly a limitation, again on smaller redshifts, when the role of non-linearities in the LSS is more prominent and this is also made manifest by the fact that BiHalofit enhances by almost $10\%$ the amplitude of obtained with Halofit at $\rho=5$ Mpc/h. Both these shortcomings are less serious at higher scales since i) we weight more the smaller angular scales, where the Limber approximation is more accurate, and ii) we move the peak of the lensing kernel to higher redshifts, where the impact of non-linearities is still relevant but attenuated.

Besides all the numerical limitations due to the usage of approximations, a take-home message that emerges from this study is that $\kappa^{LSS}_3$ is always dominant with respect to $\kappa^Q_3$ and $\kappa^{PB}_3$, regardless of the choice of the smoothing scale $\rho$ and the use of linear or non-linear matter power spectrum, as one can see from a direct comparison of Figs.~\ref{fig:lin} and \ref{fig:LSS_total}. Hence, the overall sign of the skewness in the investigated range of redshift due to the lensing is always negative, in accordance with the indication of the numerical simulation of \cite{Adamek:2018rru}. However, a deeper discussion about how to compare our results with \cite{Adamek:2018rru} is in order. This is the topic of the next section.

\section{Comparison with numerical simulations}
\label{sec:num_sim}
In this section, we will address the challenges that arise when attempting to compare our results with those obtained from ray-tracing across the numerical simulation presented in \cite{Adamek:2018rru}. We will also outline the recommended approach to ensure a valid and meaningful comparison. To begin, we provide a brief overview of the key elements employed in \cite{Adamek:2018rru}.
The ray-tracing of light-like geodesic in \cite{Adamek:2018rru} has been performed through a relativistic Universe simulated with {\it gevolution} \cite{Adamek:2015eda}, with the following cosmological parameters: the reduced Hubble constant $h = 0.67556$, the cold dark matter density $\Omega_c = 0.2638$, and the baryon density $\Omega_b = 0.048275$. Furthermore, the radiation density that includes massless neutrinos with $N_\text{eff} = 3.046$ and linear initial conditions are computed with CLASS \cite{Lesgourgues:2011re,Blas:2011rf,Lesgourgues:2011rg,Lesgourgues:2011rh} at redshift $z_\text{ini} = 127$, assuming a primordial power spectrum with amplitude $A_s = 2.215 \times 10^{-9}$ (at the pivot scale $0.05$ Mpc$^{-1}$) and spectral index $n_s = 0.9619$. In this regards, the skewness has been obtained with the following remarks:
\begin{enumerate}
\item{all the ray-traced light-like geodesics start from a point where a local overdensity has created an Halo;}
\item{structures evolve in a box with size 2.4 Gpc/$h$ whose grid space is 312.5 kpc/$h$. Hence, we expect that structures can be investigated roughly up to a scale in Fourier space of 3 $h$/Mpc;}
\item{the distribution of the obtained distance-redshift relation is binned in four bins of width 0.5, ranging from $z=0$ until $z=2$.}
\end{enumerate}
For the sake of completeness, we also report that the ray-tracing in \cite{Adamek:2018rru} is performed through a light-cone with a partial covering of the observed sky of 450 deg$^2$. Moreover, the simulations in \cite{Adamek:2018rru} link the density contrast to the gravitational potential beyond the Newtonian approximation through the second-order Hamiltonian constraint (see Eq.~(2.19) of \cite{Adamek:2021rot}).

With this in mind, we have then extrapolated from Fig.~2 in \cite{Adamek:2018rru} the values for the skewness in the four bins shown in Table~\ref{tab:skewness}.
\begin{table}[ht!]
\begin{center}
\begin{tabular}{||c c c||} 
 \hline
 Bin & Redshift Range & Skewness $\kappa_3$ 
from \cite{Adamek:2018rru} \\ [0.5ex] 
 \hline\hline
 1 & 0 - 0.5 & -2.27 \\ 
 \hline
 2 & 0.5 - 1 & -1.44 \\
 \hline
 3 & 1 - 1.5 & -0.72 \\ 
 \hline
 4 & 1.5 - 2 & -0.44 \\ 
 \hline
\end{tabular}
\caption{Values of the skewness for the distance-redshift relation distribution obtained in \cite{Adamek:2018rru} from the ray-tracing of photons across a $\Lambda$CDM Universe simulated with {\it gevolution}. Results have been obtained by binning the dataset in four redshift bins from 0 to 2 with bin width of 0.5.}
\label{tab:skewness}
\end{center}
\end{table}
The results of Table \ref{tab:skewness} seem quite in disagreement with what we have obtained in Sect.~\ref{sec:numerics} for the first two bins, whereas they match better at higher redshifts. However, we have to keep in mind the previous bullet points to understand how our results should (or better, could) be interpreted against \cite{Adamek:2018rru}.

First of all, bullet point 1 states that only regions in the simulated Universe where the overdensity is high enough to have created Halos are spanned by the ray-traced distance-redshift distribution. This motivates well the adoption of a number-count weighted prescriptions for the averages in Eq.~\eqref{eq:linear_number_count} since this naturally weights more regions where structures are more likely to have been created.

Moving forward with the discussion, bullet point 2 tells that no smoothing-scale procedure has been introduced in \cite{Adamek:2018rru}, beyond the grid space of the numerical simulation. Moreover, this implies that the expected results can probe scales up to the deeply non-linear regime and this is somehow problematic for the analytic investigations, since the validity of our perturbative scheme is at least questionable on those scales. However, this point could be easily overcame, since a smoothing scale procedure could be applied as well to the simulated Universe.

Finally, for what concerns bullet point 3, the bin width adopted in \cite{Adamek:2018rru} is quite large when compared to our infinitesimal bin formalism and this can introduce spurious contamination to the extrapolated skewness in two regards: first, by introducing background effects which are not at all related to the LSS and, secondly, by introducing a plethora of other relativistic effects, such as the contamination due to cross correlations with density fluctuations and redshift-space-distortion in Eq.~\eqref{eq:linear_number_count}, which could complicate the actual analysis. This point can be treated either by slicing the redshift space with more narrow bins or by developing our analytic formalism to the finite bin case. We plan to achieve the latter task in a forthcoming work.

Given all the assumptions and the intrinsic differences that are present between our analytical findings and the numerical ones of \cite{Adamek:2018rru}, we believe that it is quite remarkable that the two results share the same signature and differ only by an order unity factor. In our opinion, our results open an interesting window to resolve the previous bullet points and provide an ultimate comparison.

\section{Summary and Conclusions}
\label{sec:conc}
In this work we have provided for the first time an analytic evaluation of the skewness for the distance-redshift distribution in $\Lambda$CDM. The primary goal has been to provide an analytic scheme to compare and understand the analogous results obtained from numerical simulations \cite{Adamek:2018rru}. To this end, we have applied the covariant and gauge-invariant formalism for the light-cone averages detailed in \cite{Fanizza:2019pfp} to evaluate the higher-order moments of the distribution.

Within our framework, we have found that the second-order perturbations of the distance-redshift relation are enough to evaluate the leading-order terms of the skewness, whereas only linear-order corrections of the measure are needed. In this regard, among all the formally viable prescriptions for these averages, we have adopted the number-counts weighted one for our evaluation. This is of particular importance since it is ideally in line with the way the ray-tracing of photons across the above-mentioned simulation is done.

Furthermore, we have focused our treatment to the infinitesimal bin width for the redshift distribution. In this case, we have outlined several cancellations in the leading-order terms, see Sect.~\ref{sec:infinitesimal}. According to these, we are left only with three terms in the final formula~\eqref{eq:skewness_label} due to lensing, Post-Born corrections and higher-order gravitational potential. In particular, the latter sources the total skewness with the matter bispectrum, integrated along the line-of-sight. Given that, what has emerged from numerical investigation is that this also provides the biggest contribution among all. On the other hand, lensing and Post-Born terms exhibit a competitive effect which turns out to be positive and small when compared against the bispectrum one.

For what concerns a closer comparison with numerical simulations in \cite{Adamek:2018rru}, it turned out that our analytical estimation of the skewness shares the same (negative) amplitude. This is quite remarkable, given the number of approximation that we had to require to obtain our results (see Sect.~\ref{sec:num_sim} for a detailed discussion). From a physical viewpoint, our skewness from the bispectrum exhibited a strong dependence on small scales and this is not surprising, since it is a one-point function.
What is important to remark is that, in order to get a result independent of the UV behavior, we have introduced a coarse-graining (or smoothing) scale in real space and quantified the impact of this smoothing scale on the final amplitude. In this regard, we have shown that varying this coarse-graining scale from 20 to 10 Mpc/$h$ contributes to a 40$\%$ increasing of the total amplitude at small redshift ($z\sim 0.1$). This dependence is still important even though less severe for distant sources ($z\sim 1$). This seems to suggest that there is room for better agreement with the results of \cite{Adamek:2018rru}, since they have probed smaller scales than ours (see again Sect.~\ref{sec:num_sim}).

Given this last point, we believe that it is worth to investigate in the future our results along the following directions: include other relativistic effects that might be important at smaller redshift, take into account the finite-bin effect and have a closer comparison with numerical simulations where the coarse-graining scale could be introduced also in the analysis of the simulated data. This work is the starting point for the latter tasks to be achieved.

%%%%%%%%%%%%%%%%%%%%%%%%%%%%

\section*{Acknowledgements}
We are thankful to Julian Adamek and Ruth Durrer for their valuable comments on the final version of this manuscript, and to Eleonora Villa for discussions at the early stage of this project. The work of TS is supported by the Della Riccia foundation grant. TS acknowledges the support of the European Consortium for Astroparticle Theory in the form of an Exchange Travel Grant. TS was supported in part by the Pisa Section of the INFN under the program TAsP (Theoretical Astroparticle Physics). TS is thankful to the Theoretical Departments of the University of Geneva and CERN for the hospitality provided during the development of this project.
GF acknowledges support by the FCT under the program {\it ``Stimulus"} with the grant no. CEECIND/04399/2017/CP1387/CT0026 and through the research project with ref. number PTDC/FIS-AST/0054/2021. GF is also member of the Gruppo Nazionale per la Fisica Matematica (GNFM) of the Istituto Nazionale di Alta Matematica (INdAM).

%%%%%%%%%%%%%%%%%%%%%%%%%%%%

\appendix
\section{Fourth-order perturbations for the skewness}
\label{app:hopert}
In this appendix, we prove that Eq.~\eqref{eq:skewness} can be equivalently computed holds also when third and fourth order perturbations in the observable and the measure are considered. Hence, it is consistent to consider only second-order perturbations to evaluate the leading order skewness.

We then follow the same approach as the one adopted in Sect.~\ref{Sec1} and expand 
\bea
\frac{\Scal}{\Scal^{(0)}}&=&1+\sigma^{(1)}+\sigma^{(2)}+\sigma^{(3)}+\sigma^{(4)}\,,
\nonumber\\
d\mu&=&d\mu^{(0)}\left( 1+\mu^{(1)}+\mu^{(2)}+\mu^{(3)}+\mu^{(4)}\right)\,.
\eea
Hence, we obtain for the average $m$
\bea
m \equiv \overline{\left\langle \frac{\Scal}{\Scal^{(0)}} \right\rangle}
&=& 1+\left\{\overline{I[\sigma^{(2)}]}+\overline{I[\sigma^{(1)}\,\mu^{(1)}]}-\overline{I[\sigma^{(1)}]\,I[\mu^{(1)}]}\right\}\nonumber\\
&&+\left\{\overline{I[\sigma^{(4)}]}+\overline{I[\sigma^{(3)}\,\mu^{(1)}]}+\overline{I[\sigma^{(2)}\,\mu^{(2)}]}+\overline{I[\sigma^{(1)}\,\mu^{(3)}]}-\overline{I[\sigma^{(3)}]\,I[\mu^{(1)}]}\right.\nonumber\\
&&\left.-\overline{I[\sigma^{(2)}\,\mu^{(1)}]\,I[\mu^{(1)}]}-\overline{I[\sigma^{(1)}\,\mu^{(2)}]\,I[\mu^{(1)}]}-\overline{I[\sigma^{(2)}]\,I[\mu^{(2)}]}+\overline{I[\sigma^{(2)}]\,I[\mu^{(1)}]^{2}}\right.\nonumber\\
&&\left.-\overline{I[\sigma^{(1)}\,\mu^{(1)}]\,I[\mu^{(2)}]}+\overline{I[\sigma^{(1)}\,\mu^{(1)}]\,I[\mu^{(1)}]^{2}}-\overline{I[\sigma^{(1)}]\,I[\mu^{(3)}]}\right.\nonumber\\
&&\left.
+2\overline{I[\sigma^{(1)}]\,I[\mu^{(1)}]\,I[\mu^{(2)}]}
-\overline{I[\sigma^{(1)}]\,I[\mu^{(1)}]^{3}}-\overline{I[\mu^{(4)}]}\right\}\,.
\label{eq:mfourth}
\eea
First line manifestly reproduces Eq.~\eqref{eq:msecond} whereas the other lines take into account the next-to-leading order corrections. With Eq.~\eqref{eq:mfourth}, we evaluate the $\alpha$-th order moment of the distribution of $\Scal/\Scal^{(0)}$ up to the fourth order as
\beq
\overline{\left\langle\left(\frac{\Scal}{\Scal^{(0)}}-m\right)^\alpha\right\rangle}=\sum_{k=0}^\alpha(-1)^{\alpha-k}\binom{\alpha}{k}\,\left(1+A^{(2)}_{k}+B^{(4)}_{k}\right)\,,
\label{eq:generic-moment}
\eeq
where we have defined
\beq
A^{(2)}_{k}\equiv\alpha\,\left\{\overline{I[\sigma^{(2)}]}+\overline{I[\sigma^{(1)}\,\mu^{(1)}]}-\overline{I[\sigma^{(1)}]\,I[\mu^{(1)}]}\right\}+\frac{k}{2}\,\left(k-1\right)\,\overline{I[\sigma^{(1)2}]}\,,
\label{eq:Ak}
\eeq
and
\bea
B^{(4)}_{k}&=&\alpha\,\left\{\overline{I[\sigma^{(4)}]}+\overline{I[\sigma^{(3)}\,\mu^{(1)}]}+\overline{I[\sigma^{(2)}\,\mu^{(2)}]}+\overline{I[\sigma^{(1)}\,\mu^{(3)}]}-\overline{I[\sigma^{(3)}]\,I[\mu^{(1)}]}\right.\nonumber\\
&&\left.-\overline{I[\sigma^{(2)}\,\mu^{(1)}]\,I[\mu^{(1)}]}-\overline{I[\sigma^{(1)}\,\mu^{(2)}]\,I[\mu^{(1)}]}-\overline{I[\sigma^{(2)}]\,I[\mu^{(2)}]}+\overline{I[\sigma^{(2)}]\,I[\mu^{(1)}]^{2}}\right.\nonumber\\
&&\left.-\overline{I[\sigma^{(1)}\,\mu^{(1)}]\,I[\mu^{(2)}]}-\overline{I[\sigma^{(1)}\,\mu^{(1)}]\,I[\mu^{(1)}]^{2}}-\overline{I[\sigma^{(1)}]\,I[\mu^{(3)}]}\right.\nonumber\\
&&\left.+2\overline{I[\sigma^{(1)}]\,I[\mu^{(1)}]\,I[\mu^{(2)}]}-\overline{I[\sigma^{(1)}]\,I[\mu^{(1)}]^{3}}\right\}\nonumber\\
&&+k\left(k-1\right)\,\overline{I[\sigma^{(1)}\,\sigma^{(3)}]}+\frac{k}{2}\left(k-1\right)\,\overline{I[\sigma^{(2)2}]}+\frac{k}{2}\left(k-1\right)\left(k-2\right)\overline{I[\sigma^{(1)2}\,\sigma^{(2)}]}\nonumber\\
&&+\frac{k}{4!}\left(k-1\right)\left(k-2\right)\left(k-3\right)\,\overline{I[\sigma^{(1)4}]}+\frac{k}{6}\left(k-1\right)\left(k-2\right)\,\overline{I[\sigma^{(1)3}\,\mu^{(1)}]}\nonumber\\
&&-\frac{k}{2}\left(k-1\right)\,\overline{I[\sigma^{(1)2}\,\mu^{(1)}]\,I[\mu^{(1)}]}+k\left(k-1\right)\,\overline{I[\sigma^{(1)}\,\sigma^{(2)}\,\mu^{(1)}]}+\frac{k}{2}\left(k-1\right)\,\overline{I[\sigma^{(1)2}\,\mu^{(2)}]}\nonumber\\
&&-k\left(k-1\right)\,\overline{I[\sigma^{(1)}\,\sigma^{(2)}]\,I[\mu^{(1)}]}-\frac{k}{6}\left(k-1\right)\left(k-2\right)\,\overline{I[\sigma^{(1)3}]\,I[\mu^{(1)}]}\nonumber\\
&&-\frac{k}{2}\left(k-1\right)\,\overline{I[\sigma^{(1)2}]\,I[\mu^{(2)}]}+\frac{k}{2}\left(k-1\right)\,\overline{I[\sigma^{(1)2}]\,I[\mu^{(1)}]^{2}}+\left(k-1-\alpha\right)\,\overline{I[\mu^{(4)}]}\nonumber\\
&&+\left[\frac{\alpha}{2}\left(\alpha-k-1\right)+k\left(\alpha-k\right)\right]\,\left\{\overline{I[\sigma^{(2)}]}+\overline{I[\sigma^{(1)}\,\mu^{(1)}]}-\overline{I[\sigma^{(1)}]\,I[\mu^{(1)}]}\right\}^{2}\nonumber\\
&&+\frac{k}{2}\left(\alpha-k\right)\left(k-1\right)\,\overline{I[\sigma^{(1)2}]}\,\left\{\overline{I[\sigma^{(2)}]}+\overline{I[\sigma^{(1)}\,\mu^{(1)}]}-\overline{I[\sigma^{(1)}]\,I[\mu^{(1)}]}\right\}\,.
\label{eq:Bk}
\eea
Again, the second-order correction $A^{(2)}_k$ is consistent with Eq.~\eqref{eq:mu_alpha}, whereas $B^{(4)}_k$ are the potential next-to-leading order corrections.

Now, we make use of the following relations for the sum of the Newton binomial
\bea
&&\sum_{k=0}^\alpha(-1)^{\alpha-k}\binom{\alpha}{k}=0\,,
\nonumber\\
&&\sum_{k=0}^\alpha(-1)^{\alpha-k}\binom{\alpha}{k}k=\delta_{\alpha 1}\,,
\nonumber\\
&&\sum_{k=0}^\alpha(-1)^{\alpha-k}\binom{\alpha}{k}k^2=\delta_{\alpha 1}+2\,\delta_{\alpha 2}\,,
\nonumber\\
&&\sum_{k=0}^\alpha(-1)^{\alpha-k}\binom{\alpha}{k}k^3=\delta_{\alpha 1}+6\,\delta_{\alpha 2}+6\,\delta_{\alpha 3}\,,
\nonumber\\
&&\sum_{k=0}^\alpha(-1)^{\alpha-k}\binom{\alpha}{k}k^4=\delta_{\alpha 1}+14\,\delta_{\alpha 2}+36\,\delta_{\alpha 3}+24\,\delta_{\alpha 4}\,,
\label{eq:bin}
\eea
which tell us that for the skewness ($\alpha = 3$), only the third and fourth power of the index $k$ in the sum in Eq.~\eqref{eq:generic-moment} survive. It is already evident from Eq.~\eqref{eq:Bk} that terms with $k^3$ and $k^4$ only contain linear and second order perturbations. An explicit evaluations with the use of Eqs.~\eqref{eq:bin} for $\alpha=3$ returns Eqs.~\eqref{eq:skewness}.

To conclude, this proves that we need only corrections up to second-order for the observable and linear in the measure perturbation to compute the high-order moments. We remark that this is not ensured for the moments of $\Scal$, since the quantity $\Scal^{(0)}-m^{(0)}$ does not vanish on the background, and then the generic moment of the distribution of $\Scal$ includes high-order corrections for the observable and the measure.

\section{Analytic proofs for the skewness in Fourier space}
\label{app:analytic}
In this appendix, we provide the detailed derivations of $\mathcal{L}(r_1,r_2)$ in Eq.~\eqref{eq:2point} and $\mu^{LSS}_3$ in Eq.~\eqref{eq:bis1}.

\subsection{$\mathcal{L}(r_1,r_2)$}
We start from the Fourier modes of the linear gravitational potential $\psi(\eta,\bk)$ from Eqs.~\eqref{eq:Fourier_conv},
where we assume that $\psi(\eta,\bk)$ is a stochastic field such that
\beq
\overline{\psi(\eta,\bk)}=0
\qquad\text{and}\qquad
\overline{\psi(\eta_1,\bk_1)\psi(\eta_2,\bk_2)}=(2\pi)^3\delta_{D}\left( {\bf k}_1+{\bf k}_2 \right) P_\psi(k_1,\eta_1,\eta_2)\,.
\eeq
In this way, the 2-point function of a 2-D Laplacian in real space $\Delta_2$ is given by
\beq
\overline{\Delta_2\psi(r_1,\bn)\Delta_2\psi(r_2,\bn)}
=\int \frac{d^3k_1}{(2\pi)^3}
P_\psi(k_1,\eta_1,\eta_2)
\Delta_2e^{-i\bk_1\cdot\bn\,r_1}
\Delta_2e^{i\bk_1\cdot\bn\,r_2}\,,
\label{eq:twoDelta2}
\eeq
At this point, we expand the exponentials in the Fourier transforms in terms of the spherical harmonics as
\beq
e^{-i\bk\cdot\bn r}=
4\pi\sum_{\ell m} (-i)^\ell j_\ell(kr)Y_{\ell m}(\hat{\bk})Y^*_{\ell m}(\bn)\,,
\label{eq:expo_expand}
\eeq
and keep in mind, for later uses, the following orthogonal relations
\bea
\sum_m Y_{\ell m}(\bn)Y^*_{\ell m}(\bn)&=&\frac{2\ell +1}{4\pi}\,,
\nonumber\\
\int d^3 \bn Y_{\ell_1 m_1}(\bn)Y^*_{\ell_2 m_2}(\bn)&=&\delta_{\ell_1\ell_2}\delta_{m_1m_2}\,.
\label{eq:useful_SH}
\eea
In this way, Eq.~\eqref{eq:twoDelta2} becomes
\bea
\overline{\Delta_2\psi(r_1,\bn)\Delta_2\psi(r_2,\bn)}
&=&(4\pi)^2\int \frac{d^3k_1}{(2\pi)^3}
P_\psi(k_1,\eta_1,\eta_2)
\sum
i^{\ell_1+\ell_2}(-1)^{\ell_1}
\ell_1(\ell_1+1)\ell_2(\ell_2+1)
\nonumber\\
&&\times
j_{\ell_1}(k_1r_1)Y_{\ell_1 m_1}(\hat{\bk}_1)Y^*_{\ell_1 m_1}(\bn)
j_{\ell_2}(k_1r_2)Y^*_{\ell_2 m_2}(\hat{\bk}_1)Y_{\ell_2 m_2}(\bn)
\nonumber\\
&=&(4\pi)^2\int \frac{k^2_1dk_1}{(2\pi)^3}
P_\psi(k_1,\eta_1,\eta_2)
\sum
\ell^2_1(\ell_1+1)^2
\nonumber\\
&&\times
j_{\ell_1}(k_1r_1)j_{\ell_1}(k_1r_2)
Y^*_{\ell_1 m_1}(\bn)Y_{\ell_1 m_1}(\bn)
\nonumber\\
&=&4\pi\int \frac{k^2_1dk_1}{(2\pi)^3}
P_\psi(k_1,\eta_1,\eta_2)
\sum_{\ell_1}
\ell^2_1(\ell_1+1)^2(2\ell_1+1)
\nonumber\\
&&\times
j_{\ell_1}(k_1r_1)j_{\ell_1}(k_1r_2)\,.
\label{eq:A4}
\eea

It is now worth to focus on the sum over $\ell_1$ appearing into Eq.~\eqref{eq:A4}. To this end, we recall that the Legendre polynomials $P_\ell(\cos\theta)$ are eigenfunctions of the 2-D Laplacian with eigenvalues $-\ell(\ell+1)$ and that $P_\ell(1)=1$. Hence, we have
\bea
&&\sum_{\ell_1}(2\ell_1+1)\ell^2_1(\ell_1+1)^2j_{\ell_1}(k_1r_1)j_{\ell_1}(k_1r_2)
\nonumber\\
&=&\left[\Delta^2_2\sum_{\ell_1}(2\ell_1+1)j_{\ell_1}(k_1r_1)j_{\ell_1}(k_1r_2) P_{\ell_1}(\cos\theta)\right]_{\theta = 0}
\nonumber\\
&=&\left[ \Delta^2_2 j_0\left( k_1\sqrt{r^2_1+r^2_2-2r_1 r_2\cos\theta} \right) \right]_{\theta = 0}\,,
\eea
where last equality holds since the $0$-th order spherical Bessel functions can be written as
\beq
j_0\left( k_1\sqrt{r^2_1+r^2_2-2r_1 r_2\cos\theta} \right)=\sum_{\ell_1}(2\ell_1+1)j_{\ell_1}(k_1r_1)j_{\ell_1}(k_1r_2) P_{\ell_1}(\cos\theta)\,.
\eeq
Then, the explicit act of the 2-D Laplacians over $j_0$ returns
\bea
&&\left[ \Delta^2_2 j_0\left( k_1\sqrt{r^2_1+r^2_2-2r_1 r_2\cos\theta} \right) \right]_{\theta = 0}
\nonumber\\
&=&4\,r_1r_2\left[ r^2_1+r^2_2+4\,r_1r_2-2\,k_1^2\,r_1r_2\left( r_1-r_2 \right)^2 \right]\,\frac{\sin\left[ k_1\left( r_1-r_2 \right) \right]}{k_1\,\left( r_1-r_2 \right)^5}\nonumber\\
&&-4\,r_1r_2\left( r^2_1+r^2_2+4\,r_1r_2 \right)\frac{\cos\left[ k_1\left( r_1-r_2 \right) \right]}{\left( r_1-r_2 \right)^4}
\nonumber\\
&=&4 k_1^4 r_1 r_2 \left\{\frac{2 r_1 r_2\,j_2\left[k_1 (r_1-r_2)\right]}{k_1^2 (r_1-r_2)^2}
+\frac{(r_1-r_2)^2\,j_1\left[k_1 (r_1-r_2)\right]}{k_1^3 (r_1-r_2)^3}\right\}\,.
\label{eq:A7}
\eea
Now, thanks to the definition of the generalized Hankel transforms \eqref{eq:Hankel}, we can insert Eq.~\eqref{eq:A7} into \eqref{eq:A4} and prove then first of Eqs.~\eqref{eq:2point}. To this end, we relate the power spectrum of the gravitational potential $P_\psi(k,\eta_1,\eta_2)$ to the matter one $P(k,\eta_1,\eta_2)$ as 
\beq
P_\psi(k,\eta_1,\eta_2)=\frac{9}{4k^4}\frac{\HH^4_0}{a(\eta_1)a(\eta_2)}\Omega^2_{m0}P(k,\eta_1,\eta_2)\,,
\eeq
where we have used the Poisson equation to link the growth factor of the gravitational potential $D_\psi$ to the growth factor $D_1$ of the matter perturbations
\beq
D_\psi=-\frac{3}{2k^2}\frac{\HH^2_0}{a}\Omega_{m0}D_1\,,
\label{eq:transf_dict}
\eeq
as prescribed by the general dictionary provided in \cite{Castorina:2021xzs} for the relations among the different cosmological transfer functions.

\subsection{$\mu^{LSS}_3$}
In this appendix, we want to provide the detailed evaluation of Eq.~\eqref{eq:bis1}. To this end, we start by writing the Dirac-delta over the triangular shapes in Fourier space as
\bea
\delta_{D} \left( \bk_1 + \bk_2 + \bk_3 \right) &=& 8 \sum  \left( -i \right)^{\ell_1+\ell_2+\ell_3} \mathcal{G}_{\ell_1\ell_2\ell_3 }^{m_1 m_2 m_3} Y^*_{\ell_1 m_1} \left( \hat{\bk}_1 \right)  Y^*_{\ell_2 m_2} \left( \hat{\bk}_2 \right)  Y^*_{\ell_3 m_3} \left( \hat{\bk}_3 \right) 
\nonumber \\
&&\times\int dx\,x^2 j_{\ell_1} \left( k_1 x \right) j_{\ell_2} \left( k_2 x \right) j_{\ell_3} \left( k_3 x \right)\,,
\label{eq:dirac_expand}
\eea
where the integral over the auxiliary variable $x$ runs from 0 to $\infty$ and the sums are over the indexes $\ell_i$ and $m_i$ with $i=1,2,3$. We recall the definition of the Gaunt integral
\bea
\mathcal{G}_{\ell_1\ell_2\ell_3 }^{m_1 m_2 m_3} &\equiv& 
\int d\Omega \, Y_{\ell_1 m_1} \left( \bn \right)  Y_{\ell_2 m_2} \left( \bn \right)  Y_{\ell_3 m_3} \left( \bn \right) 
\nonumber\\
&=& \left( \begin{array}{ccc} 
\ell_1 & \ell_2 & \ell_3 \\ 0 & 0 & 0 
\end{array} \right) 
\left( \begin{array}{ccc}
\ell_1 & \ell_2 & \ell_3 \\ m_1 & m_2 & m_3 
\end{array} \right) \sqrt{\frac{\left(2 l_1 +1 \right) \left(2 l_2 +1 \right) \left(2 l_3 +1 \right)}{4\pi}}\,.
\eea
We remark that the integral over $x$ automatically satisfies the triangle inequality over $k_1$, $k_2$ and $k_3$ and then there is no need to explicitly write any triangular condition. We then compute the integrals over the angular directions $\bk_i$ in Eq.~\eqref{eq:3point}. By making use of Eq.~\eqref{eq:dirac_expand}, we then have
\bea
&&
\int
d \Omega_{\hat \bk_1} d \Omega_{\hat \bk_2} d \Omega_{\hat \bk_3}  \delta_D \left( \bk_1 + \bk_2 + \bk_3 \right)
\Delta_2 e^{i \bk_1 \cdot \bn r_1} 
\Delta_2 e^{i \bk_2 \cdot \bn r_2}  
\Delta_2 e^{i \bk_3 \cdot \bn r_3}  
\nonumber \\
&=&- 8 \left( 4 \pi \right)^3  \sum \ell_1 \left( \ell_1 +1 \right) \ell_2 \left( \ell_2 +1 \right) \ell_3 \left( \ell_3 +1 \right)  j_{\ell_1 }\left( k_1 r_1 \right) j_{\ell_2 }\left( k_2 r_2 \right) j_{\ell_3 }\left( k_3 r_3 \right) 
\nonumber  \\
&&
\times Y^*_{\ell_1 m_1 }\left( \bn \right) Y^*_{\ell_2 m_2 }\left( \bn \right) Y^*_{\ell_3 m_3 }\left( \bn \right)\mathcal{G}_{\ell_1\ell_2\ell_3 }^{m_1 m_2 m_3}
\int dx\,x^2 j_{\ell_1} \left( k_1 x \right) j_{\ell_2} \left( k_2 x \right) j_{\ell_3} \left( k_3 x \right)\,,
\label{eq:7.17}
\eea
where we made use of the orthonormality of spherical harmonics in Eqs.~\eqref{eq:useful_SH}. In order to move on with the evaluation, we notice that the we can get rid of the sum over $m_i$'s, since the following relation for the Gaunt integral holds
\bea
\sum_{m_1 m_2 m_3} \mathcal{G}_{\ell_1\ell_2\ell_3 }^{m_1 m_2 m_3} Y^*_{\ell_1 m_1 }\left( \bn \right) Y^*_{\ell_2 m_2 }\left( \bn \right) Y^*_{\ell_3 m_3 }\left( \bn \right)
= \left( \begin{array}{ccc}
\ell_1 & \ell_2 & \ell_3 \\ 0 & 0 & 0 
\end{array} \right)^2 
\frac{\left( 2\ell_1 +1 \right)\left( 2\ell_2 +1 \right)\left( 2\ell_3 +1 \right)}{\left( 4 \pi \right)^2}\,,
\nonumber\\
\label{eq:6.40}
\eea
thanks to the following properties
\bea
\sum_{m_1 m_2} \mathcal{G}_{\ell_1\ell_2\ell_3 }^{m_1 m_2 m_3} \mathcal{G}_{L \ell_1\ell_2 }^{M m_1 m_2} &=&
\left( \begin{array}{ccc}
\ell_1 & \ell_2 & \ell_3 \\ 0 & 0 & 0 
\end{array} \right)^2 \frac{\left( 2\ell_1 +1 \right)\left( 2\ell_2 +1 \right)}{4 \pi} \delta_{\ell_3 L} \delta_{m_3 M}\,,
\nonumber\\
Y_{\ell_1 m_1 } \left( \bn \right) Y_{\ell_2 m_2 } \left( \bn \right) &=& \sum_{L M} \mathcal{G}_{\ell_1\ell_2 L }^{m_1 m_2 M} Y^*_{L M} \left( \bn \right)\,.
\eea
Hence, by inserting Eq.~\eqref{eq:6.40} into Eq.~\eqref{eq:7.17}, we have
\bea
&&
\int
d \Omega_{\hat \bk_1} d \Omega_{\hat \bk_2} d \Omega_{\hat \bk_3}  \delta_D \left( \bk_1 + \bk_2 + \bk_3 \right)
\Delta_2 e^{i \bk_1 \cdot \bn r_1} 
\Delta_2 e^{i \bk_2 \cdot \bn r_2}  
\Delta_2 e^{i \bk_3 \cdot \bn r_3}  
\nonumber \\
&=&- 8 \left( 4 \pi \right)^3  \sum \ell_1 \left( \ell_1 +1 \right) \ell_2 \left( \ell_2 +1 \right) \ell_3 \left( \ell_3 +1 \right)  j_{\ell_1 }\left( k_1 r_1 \right) j_{\ell_2 }\left( k_2 r_2 \right) j_{\ell_3 }\left( k_3 r_3 \right) 
\nonumber  \\
&&
\left( \begin{array}{ccc}
\ell_1 & \ell_2 & \ell_3 \\ 0 & 0 & 0 
\end{array} \right)^2 
\frac{\left( 2\ell_1 +1 \right)\left( 2\ell_2 +1 \right)\left( 2\ell_3 +1 \right)}{\left( 4 \pi \right)^2}
\int dx\,x^2 j_{\ell_1} \left( k_1 x \right) j_{\ell_2} \left( k_2 x \right) j_{\ell_3} \left( k_3 x \right)\,.
\eea
Then, by using this last equation into Eq.~\eqref{eq:3point}, we finally obtain the desired result of Eq.~\eqref{eq:bis1}.

%%%%%%%%%%%%%%%%%%%%%%%%%%%%%%

\bibliography{Skewness}
\bibliographystyle{JHEP}

\end{document}